\documentclass[authoryear]{elsarticle}
\usepackage[english]{babel}
\usepackage[utf8]{inputenc}
\usepackage{amsmath}
\usepackage{amsthm}
\usepackage{amsfonts,amssymb,amsthm,amsmath}
\usepackage{graphicx}
\usepackage{xcolor}
\usepackage[colorinlistoftodos]{todonotes}
\usepackage{setspace}
\usepackage{graphicx}
\usepackage{subfig}

\newtheorem{lemma}{Lemma}

\newtheorem{remark}{Remark}
\newtheorem{theorem}{Theorem}

\usepackage{float}

\newcommand{\enegrita}{\mathbf{e}}

\newcommand{\pinegrita}{\mbox{\boldmath$\pi$}}

\newcommand{\bmap}{\emph{BMAP}}

\newcommand{\be}{\begin{eqnarray}}
\newcommand{\ee}{\end{eqnarray}}
\newcommand{\beq}{\begin{equation}}
\newcommand{\eeq}{\end{equation}}
\newcommand{\ea}{\end{eqnarray*}}
\newcommand{\ba}{\begin{eqnarray*}}
\newcommand{\edi}{\end{displaymath}}
\newcommand{\bdi}{\begin{displaymath}}

\newcommand{\map}{\emph{MAP}}
\newcommand{\mmpp}{\emph{MMPP}}
\newcommand{\bmmpp}{\emph{BMMPP}}

\usepackage{colortbl}

\definecolor{maroon}{cmyk}{0,0.87,0.68,0.32}

\usepackage[detect-all]{siunitx}

\usepackage{multirow}

\begin{document}
 
 \title{\textcolor{black}{Fitting procedure for the two-state} Batch Markov modulated Poisson process}

\author[uc3m]{Yoel G. Yera\corref{cor1}}
\ead{yoelgustavo.yera@uc3m.es}
\author[uc3m,UC3M-BS]{Rosa E. Lillo}
\ead{rosaelvira.lillo@uc3m.es}
\author[uca,imus]{Pepa Ramírez-Cobo}
\ead{pepa.ramirez@uca.es}

\cortext[cor1]{Corresponding author}

\address[uc3m]{Department of Statistics, Universidad Carlos III de Madrid, Spain}
\address[UC3M-BS]{UC3M-Santander Big Data Institute, Madrid, Spain}
\address[uca]{Department of Statistics and Operations Research, Universidad de C\'adiz, Spain}

\address[imus]{IMUS, Institute of Mathematics Universidad de Sevilla, Spain}

\providecommand{\keywords}[1]{\textbf{\textit{Keywords: }} #1}

\begin{abstract}
The Batch Markov Modulated Poisson Process ($\bmmpp$) is a subclass of the versatile Batch Markovian Arrival process ($\bmap$)  which has  been proposed for the modeling of dependent events occurring in batches (as group arrivals, failures or risk events). This paper focuses on exploring the possibilities of the $\bmmpp$ for the modeling of real phenomena involving point processes with group arrivals. The first result in this sense is the characterization of the two-state $\bmmpp$ with maximum batch size equal to $K$, the $\bmmpp_2(K)$, by a set of moments related to the inter-event time and batch size distributions. This characterization leads to a sequential fitting approach via a moments matching method. \textcolor{black}{The performance of the novel fitting approach is illustrated on both simulated and a real teletraffic data set, and compared to that of the EM algorithm. In addition, as an extension of the inference approach, the queue length distributions at departures in the queueing system $\bmmpp/M/1$ is also estimated.}
\end{abstract}

\maketitle

\keywords{\textcolor{black}{stochastic processes}, Markov modulated Poisson process ($\mmpp$), Batch Markovian arrival process ($\bmap$), Identifiability, Moments matching method, teletraffic data, \textcolor{black}{$\bmap/G/1$ queueing system.}}

\section{Introduction}
In this work, we propose a fitting approach \textcolor{black}{for correlated times} between the occurrence of events (that may occur in batches) via a general subclass of the Batch Markovian Arrival Process ($\bmap$), {the Batch Markov Modulated Poisson Process ($\bmmpp$)}. Events can be understood from multiple contexts: failures in an electronic system, arrivals of packets of bytes in a teletraffic setting, arrivals of customers in a queue or risk events, among others. The $\bmap$ constitutes a large class of point processes that allows for non-exponential and dependent times between the occurrence of events, which may occur in batches (that is, more than one event at a time). $\bmap$s were first introduced by \cite{Neu79}, although the current and more tractable description is due to \cite{Luca91}. It is known that stationary $\bmap$s are capable of approximating any stationary batch point process \citep{Asm93} which points out the versatility of the process. In addition, the $\bmap$ is a tractable process from an analytical viewpoint, since most of the associated descriptors and probabilities of interest can be computed in a straightforward way. For these reasons, $\bmap$s have been widely considered in a number of real-life contexts, as queueing, teletraffic, reliability, hydrology or insurance, where dependent events (possibly occurring in batches) are commonly observed. For a recent account of the literature on $\bmap$s applications, we refer the reader to  \cite{Ram142}, \cite{Ban15}, \cite{Liu15}, \cite{Mon15}, \cite{Singh16}, \cite{Sik16}, \cite{Ban17}, \cite{Gho17},  and  \cite{Buch17}.

{The complexity and versatility of $\bmap$s increase with the number of parameters defining the process, which is related to the identifiability issue. In the context of $\bmap$s, the lack of identifiability may be formulated along the lines of \cite{Ryd96} or \cite{Ram10}. Specifically, if  $T_{n}$ and $B_{n}$ represent the time between the $(n-1)$-th and $n$-th events ocurrences, and the batch size of the $n$-th event in a $\bmap$ noted by $\mathcal{B}$, then $\mathcal{B}$ is said to be non-identifiable if there exists a differently parametrized $\bmap$, noted as $\widetilde{\mathcal{B}}$, such that 
\begin{equation*}
\label{def: nonidentifiability}
\left(T_{1},\ldots,T_{n},B_{1},\ldots,B_{n}\right)\overset{d}{=}\left(\widetilde{T}_{1},\ldots,\widetilde{T}_{n},\widetilde{B}_{1},\ldots,\widetilde{B}_{n}\right) \quad \text{for all $n\geq1$},
\end{equation*}
where $\overset{d}{=}$ denotes equality of joint distributions, and $\widetilde{T}_{n}$ and $\widetilde{B}_{n}$ represents the inter-event times and batch sizes of the $\bmap$ noted as $\widetilde{\mathcal{B}}$. } \textcolor{black}{The lack of a unique representation affects to the statistical inference of the process: if the process is non-identifiable this means that the likelihood function of the inter-event times will be multimodal, and therefore any likelihood-based fitting algorithm will turn out strongly dependent on the starting point.  Because of this, the issue of identifiability has been broadly studied in the literature for certain classes of $\bmap$s, see for instance \cite{GreenDA,Bean,He1, He2, He3,Ryd96,Ram10,Ram12,Rod162,Rod16,Yera}. As a result, it is known that both the Markov modulated Poisson process ($\mmpp$) \citep{Heffes.packetized,Sco99,Sco03,Fea06,Lan13} and its batch counterpart, the $\bmmpp$ considered in this paper, are identifiable.}

Taking advantage of the identifiability of the $\bmmpp$, this paper addresses the problem of statistical inference for the two-state Batch Markov modulated Poisson process, noted  $\bmmpp_2(K)$ where $K$ represents the maximum batch size. \textcolor{black}{The choice of the $\bmmpp_2(K)$ over higher order $\bmmpp_m(K)$s (that is, processes with $m\geq 3)$ is motivated by some reasons. First, the considered model is characterized by a smaller number of parameters, a fact that eases the estimation process. Second, it is expected that higher order $\bmmpp_m(K)$s present more versatility and be able to model more complex patterns (\cite{Rod161} gives some empirical results in this line); however, up to our knowledge there are no studies exploring in depth such degrees of versatility and in consequence, it is impossible to know \emph{a priori} which is the smallest order $m$ that shall be needed for fitting a given data set. Finally, as will be shown in Section \ref{momcharact}, the $\bmmpp_2(K)$ can be completely characterized in terms of a set of $2(K+1)$ moments related to the inter-event times and batch sizes distribution, which naturally leads to a moments-matching fitting approach. However, this characterization in terms of moments remains as an open question for the case $m \geq 3$, and will be the subject of future work as indicated in the conclusions section.}

\textcolor{black}{The contribution of this paper is two-fold. On one hand and as commented before, it is proven that the $\bmmpp_2(K)$, which is represented by $2(K+1)$ parameters, is characterized by a set of $2(K+1)$ moments concerning the distributions of both the inter-event times and batch sizes. On the other hand, a sequential estimation approach for fitting real data sets is derived and illustrated for simulated and real data sets. At this point, some remarks concerning statistical inference for the $\bmap$s need to be made. First, concerning the observed information, in most of papers it is assumed that the sequence of inter-event times, $\boldsymbol{t} = (t_1, t_2, . . . , t_n)$ (and if it is the case, of batch sizes  $\boldsymbol{b} = (b_1, b_2, . . . , b_n)$) constitute the available observed samples. This implies that many components of the process (as the transition times or sequence of visited states) remain unobserved. Other authors instead consider that the observed information is related to the counting process (number of accumulated events at some time instants, for example), see \cite{andersen2002use,arts2017multi,nasr2018map}. In this paper, the considered approach will be the first one. Second, there are a number of papers in the literature dealing with strategies (either Bayesian, frequentist or moments matching based) for estimation of some types of $\map$s (characterized by single events at a time). In these works, either the considered $\map$s are identifiable (as the $\mmpp$, see \cite{Ryden94,ryden961,Sco99}) or non-identifiable but, with a known canonical form (as the $\map_2$ see \cite{Eum,Bod08,Carpepa,BayesPepa}). If events occurring in batches are however observed, fewer works dealing with inference for the $\bmap$ can be found and up to our knowledge they all are based on the EM algorithm, see for example \cite{Breuer,Klemm}. In this paper the performance of the proposed sequential fitting algorithm shall be compared to that of the EM, as designed in such papers.}

{The paper is structured as follows. After a brief review of the 
$\bmmpp_2(K)$ in Section \ref{Sec:DescBMMPP2}, the moments characterization for the $\bmmpp_2(K)$ is proven in Section \ref{momcharact}. \textcolor{black}{Section \ref{caseK2} analyzes in depth the case $K=2$ and Section \ref{K3} extends the findings for the case $K\geq 3$. The characterization in terms of moments leads to the sequential fitting algorithm presented and illustrated in Section \ref{fitting}. After the detailed description of method in Section \ref{fitting2}, its performance on simulated traces is illustrated in Section \ref{simu} and a comparison with the EM algorithm is provided in Section \ref{EM}. Finally, Section \ref{teletraffic} considers a real application of the novel approach: the modeling of a well-referenced data set from the teletraffic context. In the numerical analyses, the estimation of the queue length distribution at departures in a  $\bmmpp/M/1$ queueing system is also considered.} Finally, Section \ref{conclusions} presents conclusions and delineates possible directions for future research.

\section{Description of the stationary $\bmmpp_2(K)$}\label{Sec:DescBMMPP2}

In this section, the two-state Batch Markov modulated Poisson process, noted $\bmmpp_2(K)$, where $K$ is the maximum batch size, is formally defined. Also, some
properties that will be used throughout this paper are reviewed. \textcolor{black}{Consider a two-state Markov process $J(t)$ with generator $Q$ on $\{1,2\}$. For each state $i\in\{1,2\}$, events occur according a Poisson process with rate $\lambda_i$ and each event has a batch distribution on $\{1,\ldots, K\}$ that also depends on $J(t)$. In other words, whenever $J(t) = i$, it is said that the process is in state $i$ at time $t$ and this status remains unchanged while the process remains in this state. As soon as the Markov process enters another state $j$ ($j\in\{1,2\}$), then the Poisson process alters accordingly.} Specifically, the $\bmmpp_2(K)$ behaves as follows:  at the end of an exponentially distributed sojourn time in state $i$, with mean $1/\lambda_i$, two possible state transitions can occur. First, with probability $p_{ij0}$, no event occurs and the system enters into a different state $j \neq i$. Second, with probability \textcolor{black}{$p_{iik}$}},  an event of batch size $k$ is produced if the state of the process is $i$, and the system continues in the same state. It is clear that
 
$$ p_{ij0}+\sum_{k=1}^Kp_{iik}=1 \qquad i,j=1,2, \quad i\neq j.$$

\textcolor{black}{A $\bmmpp_2(K)$ can be thus expressed in terms of the initial probability vector and the parameters $\{\boldsymbol{\lambda},\boldsymbol{P_0},..,\boldsymbol{P_K}\}$, where $\boldsymbol{\lambda}=(\lambda_1, \lambda_2)$, and $\boldsymbol{P_0}$, $\boldsymbol{P_1}$,...,$\boldsymbol{P_K}$ are $2 \times 2$ transition probability matrices with $(i,j)$-th elements $p_{ijk}$, for $k=1,...,K$. On the other hand, instead of transition probability matrices, any $\bmmpp_2(K)$ can also be characterized in terms of rate (or intensity) matrices. In the case of the $\bmmpp_2(K)$, these rate matrices are $\{\boldsymbol{D_0},\boldsymbol{D_1},...,\boldsymbol{D_K}\}$ where}

\begin{eqnarray}\nonumber
\boldsymbol{D_0}&=& \left( \begin{array}{cc}
x & y  \\
r & u  \end{array} \right)\\ \label{representation}
\boldsymbol{D_k}&=&\left( \begin{array}{cc}
w_k & 0  \\
0 & q_k  \end{array} \right),\quad 1\leq k \leq K-1\\ \nonumber
\boldsymbol{D_K}&=&\left( \begin{array}{cc}
-x -y -\sum_{i=1}^{K-1}w_i & 0  \\
0 & -r-u-\sum_{i=1}^{K-1}q_i  \end{array} \right).
\end{eqnarray}
\textcolor{black}{Under this representation, the transitions where no event occurs are governed by the $\boldsymbol{D_0}$, while the transitions characterized by a batch event of size $k$ are governed by $\boldsymbol{D_k}$. In addition, the definition of the rate matrices implies that
$\boldsymbol{Q}=\sum_{k=0}^K\boldsymbol{D_k}$
is the infinitesimal generator of the underlying Markov process $J(t)$, with stationary probability vector {$\boldsymbol{\pi} = (\pi^* , 1-\pi^*)$}, satisfying $\boldsymbol{\pi} \boldsymbol{Q} = 0$ and $\boldsymbol{\pi} \boldsymbol{e} = 1$, where $\boldsymbol{e}$ is a {column} vector of ones. The relationship between the transition probabilities matrices representation and the one based of rate matrices is
$$ x = -\lambda_1,\ u = -\lambda_2,\ y = \lambda_1 p_{120},\ r = \lambda_2 p_{210},\ w_k = \lambda_1 p_{11k},\ q_k = \lambda_2 p_{22k}. $$ In this work the characterization given by (\ref{representation}) will be the considered from now on.}

\begin{figure}[h!]
\centering
\includegraphics[scale = .5]{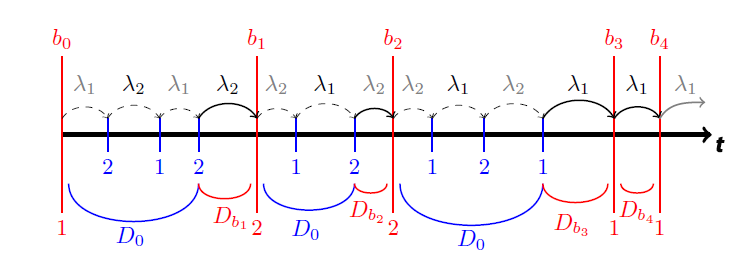}\\
\caption{Transition diagram for the $\bmmpp_2(K)$. The dashed line corresponds to transitions without events, governed by $\boldsymbol{D_0}$, and the solid lines correspond to transitions of size $b_k$, governed by $\boldsymbol{D_{b_k}}$.}\label{fig:BMMPP2K}
\end{figure}

For a better understanding of the considered process, Figure \ref{fig:BMMPP2K}  illustrates a realization of the $\bmmpp_2(K)$, where the dashed line corresponds to transitions where no events occur, and the solid lines correspond to transitions where an event of size $b_i\in\{1,\ldots, K\}$ occurs.  

It is important to note that if $\mathcal{B}_K=\left\{\boldsymbol{D_0},\ldots, \boldsymbol{D_K}\right\}$ represents a $\bmmpp$ with maximum batch size equal to $K$, then  $\mathcal{M}=\left\{\boldsymbol{G_0}=\boldsymbol{D_0},\ \boldsymbol{G_1}=\boldsymbol{D_1}+...+ \boldsymbol{D_K}\right\}$ defines a Markov modulated Poisson process ($\mmpp$), which satisfies the same inter-event times properties as $\mathcal{B}_K$, but is not able to model events occurring in batches.

\textcolor{black}{
\begin{remark}\label{rosa}
Some authors define the $\bmmpp$ taking $p_{iik}=p_{jjk}$ for all $i\neq j$, see for example \cite{Chakravarthy}. In this case, the intensity matrices are expressed as  $\boldsymbol{D_0} = \boldsymbol{Q}- \boldsymbol{\Delta}(\boldsymbol{\delta})$, where $\boldsymbol{Q}$ is the infinitesimal generator of the underlying Markov process $J(t)$ and $\boldsymbol{\Delta}(\boldsymbol{\delta})$ is a non-negative diagonal matrix; and $\boldsymbol{D_k} = \boldsymbol{\Delta}(\boldsymbol{\delta})\boldsymbol{\Delta}(p_k)$, for all $k \geq 1$, where $\boldsymbol{\Delta}(p_k)$ is a non-negative diagonal matrix with $i^{th}$ diagonal entry given by $p_k$, being $\sum_{k=1}^K p_k=1$. This is a particular case of the process introduced by  \cite{Luca93} and denoted as a $\map$ with i.i.d. batch arrivals. These processes have  the advantage of being simpler than the one considered in this paper, but alse have the drawback that $corr(T,B)$ and $\rho_B(1)$ are both null by construction. 
(See the supplementary material for a proof of these properties). As will be seen in Section \ref{simu} , using the simple model in the estimation with data leads to a worse performance in modeling and its posterior use. 
\end{remark}}

\subsection{Performance measures regarding the inter-event times and batch sizes}
A review of the performance measures concerning the inter-event times and batch sizes in a $\bmmpp_2(K)$ is given next. If $S_n$ denotes the state of the underlying Markov process at the time of the $n$-th event, $B_n$ the batch size of that event and  $T_n$ the time between the $(n-1)$-th and $n$-th events, then the process { $\{S_{n-1},\sum_{i=1}^n \textcolor{black}{T_i}, B_n\}_{n=1}^\infty$},  is a Markov renewal process (see for example, \cite{Cha10}). Furthermore, if 
$$\boldsymbol{D} =\sum_{k=1}^K \boldsymbol{D_k},$$
then $\{S_n\}_{n=0}^\infty$ is a Markov chain with transition matrix 
\begin{equation*}\label{Pstar}
\boldsymbol{P^*} = (-\boldsymbol{D_0})^{-1}\boldsymbol{D}.
\end{equation*}

On the other hand, the variables $T_n$s are phase-type distributed  with representation $\{\boldsymbol{\phi},\boldsymbol{D_0}\}$, where $\boldsymbol{\phi}$ is the stationary probability vector associated to $\boldsymbol{P^*}$ computed as $\boldsymbol{\phi}=(\boldsymbol{\pi} \boldsymbol{D} \boldsymbol{e})^{-1}\boldsymbol{\pi} \boldsymbol{D}$ (see \cite{Lat90} and \cite{Cha10}). In consequence, the moments of  $T_n$ in the stationary case are given by
\begin{equation}\label{eq:MAP-mu}
\mu_r = E (T^r ) = r!\boldsymbol{\phi} (-\boldsymbol{D_0})^{-r}\boldsymbol{e}, \quad \textrm{for } r\geq 1,
\end{equation}
and the auto-correlation function of the sequence of inter-event times is 
\begin{equation}\label{eq:Tcor}
\rho_T(l)=\rho (T_1, T_{l+1}) = \gamma^l \frac{\mu_2-2\mu_1^2}{2(\mu_2-\mu_1^2)},\quad \text{for }l>0. 
\end{equation}
In (\ref{eq:Tcor}), $\gamma$ is one of the two eigenvalues of the transition matrix $\boldsymbol{P^*}$ (as $\boldsymbol{P^*}$ is stochastic, then necessarily the other eigenvalue is equal to $1$). According to \cite{Kang95}, the value of $\gamma$ in (\ref{eq:Tcor}) is non-negative in the case of the $\bmmpp_2$ and $\mmpp_2$ which implies that the inter-event times are always positively correlated.

Also, from \cite{Rod16}, the mass probability function of the stationary batch size, $B$, is
$$P(B = k) = \boldsymbol{\phi}(-\boldsymbol{D_0})^{-1}\boldsymbol{D}^k\boldsymbol{e}, \quad \textrm{for }  k = 1, . . . , K,$$
from which the moments of $B$ are obtained as
\begin{equation}\label{Eq:BMoments}
\beta_r=E[B^r] = \boldsymbol{\phi}(-\boldsymbol{D_0})^{-1}\boldsymbol{D_r^*}\boldsymbol{e}, \quad \textrm{for } r\geq 1,
\end{equation}
where 
$\boldsymbol{D_r^*}= \sum_{k=1}^K k^r\boldsymbol{D_k}$. Also, the autocorrelation function in the stationary version of the process $\rho_B(l)$ is given by
\begin{equation*}\label{eq:Bcor}
\rho_B(l)=\rho(B_1,B_{l+1}) =\frac{ \boldsymbol{\phi}(-\boldsymbol{D_0})^{-1}\boldsymbol{D_1^*}[(-\boldsymbol{D_0})^{-1}\boldsymbol{D}]^{l-1}(-\boldsymbol{D_0})^{-1}\boldsymbol{D_1^*}\boldsymbol{e} - \beta_1^2}{\sigma_B^2},
\end{equation*}
where $\beta_1$ and $\sigma_B^2 = \beta_2-\beta_1^2$ are  computed from (\ref{Eq:BMoments}). 

\textcolor{black}{
Using the Laplace-Stieltjes transform (LST) of the $n$ first inter-event times and batch sizes 
of a stationary $\bmap_2(K)$ given in \cite{Rod16}, then $E[TB]$ is found as
\begin{equation}\label{joint_moment}
\eta=E[TB] = \phi(-\boldsymbol{D_0})^{-2}\boldsymbol{D_1^*}\boldsymbol{e}.
\end{equation}
See the supplementary material for a proof. From this, the covariance between $T$ and $B$ is obtained as
\begin{equation*}
cov(T,\ B)=\phi(-\boldsymbol{D_0})^{-2}\boldsymbol{D_1^*}\boldsymbol{e}-\boldsymbol{\phi}(-\boldsymbol{D_0})^{-1}\boldsymbol{e}\boldsymbol{\phi}(-\boldsymbol{D_0})^{-1}\boldsymbol{D_{1}^*}\boldsymbol{e}.
\end{equation*}}

\subsection{Performance measures regarding the counting process}\label{counting}

Consider a stationary $\bmmpp_2(K)$ represented by $\mathcal{B}_K=\{\boldsymbol{D_{0}}, \boldsymbol{D_{1}},\dots,\boldsymbol{D_K}\}$ with underlying phase process $\{J(t)\}_{{t\geq 0}}$. Then, the counting process $N(t)$ represents the number of events {that occur} in $(0,t]$. For $n \in \mathbb{N}$ and $t\geq 0$, let $\boldsymbol{P}(n,t)$ denote the $2\times 2$ matrix whose $(i,j)$-th element is 
\begin{equation*}\label{Pij}
\boldsymbol{P}_{ij}(n,t)=P\left(N(t)=n,\ J(t)=j \mid N(0)=0,\ J(0)=i\right),
\end{equation*}
for $1\leq i,j \leq 2$. From the previous definition it is clear that
\begin{equation}\label{PN}
p(n,t)=P\left(N(t)=n\mid N(0)=0\right)=\pinegrita \boldsymbol{P}(n,t) \enegrita.
\end{equation}
The values of the matrices $\boldsymbol{P}(n,t)$ cannot be computed in closed-form. However, their numerical computation is straightforward from the \emph{uniformization method} addressed in \cite{NeutsLi}. 

If the interest is focused on counting the events of a specific size $k\in\{1,\ldots,K\}$, define $N(t,k)$ as the number of such events that have occurred up to time $t$. Then, it is clear that $N(t,k)\overset{d}{=}N_{\mathcal{M}}^k(t)$, where $N_{\mathcal{M}}^k(t)$ is the counting process of the $\mmpp$ given by $\mathcal{M}=\{\boldsymbol{G_0}=\boldsymbol{D_0}+\boldsymbol{D_1}+\ldots+\boldsymbol{D_{k-1}}+\boldsymbol{D_{k+1}}+\ldots+\boldsymbol{D_K},\ \boldsymbol{G_1}=\boldsymbol{D_k}\}$. Therefore, the probabilities of $N(t,k)$ can be computed as those of $N_{\mathcal{M}}^k(t)$, via expression (\ref{PN}).

Some moments concerning the counting process are as follows, see \cite{narayana} or \cite{Eum}. In the stationary version of the process, the mean number of events in an interval of length $t$ (known as the \emph{Palm function}) is
\begin{equation*}\label{expected_losses}
E\left[N(t)\right]=\lambda^{\star}t,
\end{equation*}
where $\lambda^{\star}=\mu_1^{-1}$ represents the events rate. The variance of that count is given by
\begin{equation}\label{variance_count}
V\left[ N(t)\right] = (1+2\lambda^{\star})E\left[N(t)\right]-2\pinegrita D\left(\enegrita\pinegrita +\boldsymbol{Q}  \right)^{-1}\boldsymbol{D}\enegrita t-2\pinegrita \boldsymbol{D}\left(\boldsymbol{I}-e^{\boldsymbol{Q}t}  \right)\left(\enegrita \pinegrita+\boldsymbol{Q}\right)^{-2}\boldsymbol{D} \enegrita.
\end{equation}

\section{Moments characterization}\label{momcharact}
In this section we prove that the $\bmmpp_2(K)$ is completely characterized by a set of $2(K+1)$ moments. As will be seen, the results are based {on} \cite{Bod08}, which provide a canonical representation for the two-state $\map$. The case where $K=2$ shall be first addressed to later consider the generalization for an arbitrary batch size.

\subsection{The $\mmpp_2$, the $\map_2$ and their canonical representations}\label{mmppmap}
As previously commented, in this work we deal with the $\bmmpp_2(K)$ which is the batch counterpart of the well-known $\mmpp_2$. As described in Section 1, the $\mmpp_2$ is an identifiable subclass of $\map_2$, a general, non-identifiable point process that includes both renewal processes (phase type renewal processes as the Erlang and hyperexponential renewal process) and non-renewal processes, as is the case of the $\mmpp_2$. It is common in the literature to represent the $\map_2$ by the rate matrices
\begin{equation}\label{map2}
\boldsymbol{G_0}=\left(\begin{array}{cc}
x & y  \\
r & u  \end{array} \right), \qquad \boldsymbol{G_1}=\left(\begin{array}{cc}
w & -x-y-w  \\
 v & -r-u-v \end{array} \right),
 \end{equation}
where $\{x,y,r,u,w,v\}$ are defined in similar way as in (\ref{representation}) (see \cite{Ram10} for more details). Then, a $\mmpp_2$ will be defined as (\ref{map2}) such that $w=-x-y$ and $v=0$,
\begin{equation}\label{mmpp2}
\boldsymbol{G_0}=\left(\begin{array}{cc}
x & y  \\
r & u  \end{array} \right), \qquad \boldsymbol{G_1}=\left(\begin{array}{cc}
-x-y & 0  \\
 0 & -r-u \end{array} \right).
\end{equation}
Without loss of generality, we will assume from now on that $x+y\geq r+u$ (otherwise, an equivalent process is obtained by permuting the states).  Note that representation (\ref{mmpp2}) implies that events only occur at self-transitions of the underlying Markov chain, and every self-transition produces an event. 

Even though the $\map_2$ in (\ref{map2}) is non-identifiable, \cite{Bod08} {provide} a canonical, unique, representation; in particular, if $\gamma>0$ (see Eq. \ref{eq:Tcor}), as it is the case of the $\bmmpp_2(K)$ and $\mmpp_2$, then, the canonical form of (\ref{map2}) is given by
\begin{equation}\label{Canform1}
\boldsymbol{G_0^c}=\left(\begin{array}{cc}
-\zeta_1 & (1-a)\zeta_1  \\
0 & -\zeta_2  \end{array} \right), \qquad \boldsymbol{G_1^c}=\left(\begin{array}{cc}
a\zeta_1 & 0  \\
 (1-b)\zeta_2 & b\zeta_2 \end{array} \right),
\end{equation}
for certain exponential rates $\zeta_1$, $\zeta_2$ and probabilities $a$ and $b$. The canonical form implies that all equivalent representations of a $\map_2$ as in (\ref{map2}) with associated $\gamma$ satisfying $\gamma>0$, can be written - in unique way - as in (\ref{Canform1}). \cite{Bod08} also {show} that any $\map_2$ as in (\ref{map2}) can be completely characterized by four moments regarding the inter-event time distribution, namely, the first, second and third moment of the inter-event time distribution, $\mu_1,\ \mu_2,\ \mu_3$, and the first-lag autocorrelation coefficient of the inter-event times, $\rho_T(1)$, see Eq. (\ref{eq:MAP-mu}) and Eq. (\ref{eq:Tcor}), for their explicit expressions. Indeed, there exists a one-to-one correspondence between the parameters of the canonical form (\ref{Canform1}) $\{\zeta_1,\zeta_2,a,b\}$ and the moments $\{\mu_1,\mu_2, \mu_3, \rho_T(1)\}$. As will be seen in the next sections, this fact shall be the base for proving the characterization of the $\bmmpp_2(K)$ in terms of a set of moments. However, in order to find such characterization, the canonical form as in (\ref{Canform1}) for a $\mmpp_2$ given by (\ref{mmpp2}) needs to be found. The following result provides such canonical representation.

\begin{lemma}\label{FromMMPPToCanForm}
Let $ \mathcal{G} = \left\{\boldsymbol{G_0},\boldsymbol{G_1}\right\}$  represent a $\mmpp_2$ as in (\ref{mmpp2}).  Then representation $\mathcal{G}$ is equivalent to the canonical representation $ \mathcal{G}^c = \left\{\boldsymbol{G_0^c},\boldsymbol{G_1^c}\right\}$, where
\begin{eqnarray*}
\boldsymbol{G_0^c}&=&\left( \begin{array}{cc}x - y\dfrac{x-2r - u - \sqrt{(u-x)^2 + 4ry}}{x + 2y - u + \sqrt{(u-x)^2 + 4ry}} & -y\dfrac{2(r + u - x - y)}{x + 2y - u  + \sqrt{(u-x)^2 + 4ry}}\\
0 & u+y\dfrac{ x - 2r - u - \sqrt{(u - x)^2 + 4ry}}{x + 2y - u + \sqrt{(u-x)^2 + 4ry}}\end{array} \right),\\
\boldsymbol{G_1^c}&=&\left( \begin{array}{cc} -x-y & 0 \\
-\dfrac{x-2r-u-\sqrt{(u-x)^2+4yr}}{2} & -r-u \end{array} \right).
\end{eqnarray*}
\end{lemma}

\begin{proof}
The proof follows \cite{Rod162}, where a similarity transform via an invertible matrix $\boldsymbol{A}$ satisfying $\boldsymbol{A}\enegrita = \enegrita$,
\begin{equation}\label{similarity}
\boldsymbol{G_0^c} = \boldsymbol{A} \boldsymbol{G_0} \boldsymbol{A}^{-1},\ \boldsymbol{G_1^c} = \boldsymbol{A} \boldsymbol{G_1} \boldsymbol{A}^{-1}
\end{equation}
is used to convert any $\map_2$ as in (\ref{map2}) in its canonical form. In particular, for the $\mmpp_2$, $\boldsymbol{A}^{-1}$ can be easily found as
$$\boldsymbol{A}^{-1}=\begin{array}{l}
\left( \begin{array}{cc} 1 & 0 \\
-\dfrac{x-2r-u-\sqrt{(u-x)^2+4yr}}{x+2y-u+\sqrt{(u-x)^2+4yr}} & \dfrac{2(x+y-u-r)}{x+2y-u+\sqrt{(u-x)^2+4yr}} \end{array} \right).
\end{array} $$
Hence, from (\ref{similarity}), the result is obtained. 
\end{proof}
Lemma \ref{FromMMPPToCanForm} shows how representation (\ref{Canform1}) can be obtained from (\ref{mmpp2}). In analogous way, the opposite transformation can be found, as the following Lemma \ref{Le:FromCanFormToMMPP} shows.

\begin{lemma}\label{Le:FromCanFormToMMPP}
Let a $\mmpp_2$ be represented in canonical way by $ \mathcal{G}^c = \left\{\boldsymbol{G_0^c},\boldsymbol{G_1^c}\right\}$ as in (\ref{Canform1}). Then, its representation $ \mathcal{G} = \left\{\boldsymbol{G_0},\boldsymbol{G_1}\right\}$ as in (\ref{mmpp2}) is given by
\begin{eqnarray*}
\boldsymbol{G_0}&=&\frac{1}{a\zeta_1 - b\zeta_2}\left( \begin{array}{cc}  \zeta_1\zeta_2 - a\zeta_1^2 - a\zeta_1\zeta_2 + ab\zeta_1\zeta_2  & -a^2\zeta_1^2 + a\zeta_1^2 + \zeta_2a\zeta_1 - \zeta_2\zeta_1\\
(\zeta_1 - b\zeta_2)(\zeta_2 - b\zeta_2) & -\zeta_1\zeta_2 + b\zeta_2^2 + b\zeta_1\zeta_2 - ab\zeta_1\lambda_2\end{array} \right)\\\\
\boldsymbol{G_1}&=&diag(G_1^c).
\end{eqnarray*}
\end{lemma}
\begin{proof}
From Lemma \ref{FromMMPPToCanForm} the canonical form associated to a $\mmpp_2$ can be written as
\begin{eqnarray*}
\textcolor{black}{\boldsymbol{G_0^c}}&=&\left(\begin{array}{cc}
-\zeta_1 & (1-a)\zeta_1  \\
0 & -\zeta_2  \end{array} \right)\\
&=&\left( \begin{array}{cc}x - y\dfrac{x-2r - u - \sqrt{(u-x)^2 + 4ry}}{x + 2y - u + \sqrt{(u-x)^2 + 4ry}} & -y\dfrac{2(r + u - x - y)}{x + 2y - u  + \sqrt{(u-x)^2 + 4ry}}\\
0 & u+y\dfrac{ x - 2r - u - \sqrt{(u - x)^2 + 4ry}}{x + 2y - u + \sqrt{(u-x)^2 + 4ry}}\end{array} \right),\\
\textcolor{black}{\boldsymbol{G_1^c}}&=&\left(\begin{array}{cc}
a\zeta_1 & 0  \\
 (1-b)\zeta_2 & b\zeta_2 \end{array} \right)\\
 &=&\left( \begin{array}{cc} -x-y & 0 \\
-\dfrac{x-2r-u-\sqrt{(u-x)^2+4yr}}{2} & -r-u \end{array} \right).
\end{eqnarray*}
Solving for $x,y,r,u$, the result is obtained. The proof that $\boldsymbol{G_0}$ and $\boldsymbol{G_1}$ are well defined can be found in the supplementary material.
\end{proof}
 
\subsection{Moments characterization for the $\bmmpp_2(2)$}\label{caseK2}
Consider a $\bmmpp_2(2)$ represented by $\mathcal{B}_2=\left\{\boldsymbol{D_0},\boldsymbol{D_1},\boldsymbol{D_2} \right\}$ where, according to (\ref{representation}),
\begin{equation}\label{K2}
\boldsymbol{D_0}=\left(\begin{array}{cc}
x & y  \\
r & u  \end{array} \right), \
\boldsymbol{D_1}=\left(\begin{array}{cc}
w & 0  \\
0 & q  \end{array} \right), \
\boldsymbol{D_2}=\left(\begin{array}{cc}
-x-y-w & 0  \\
 0 & -r-u-q \end{array} \right).
 \end{equation}
Note that $\mathcal{M}=\{\boldsymbol{G_0}=\boldsymbol{D_0},\ \boldsymbol{G_1}=\boldsymbol{D_1}+\boldsymbol{D_2}\}$ is a representation of a $\mmpp_2$, and therefore, according to Lemma 1,  $\mathcal{M}$ has a canonical form as in (\ref{Canform1}). Then, from \cite{Bod08} such canonical representation can be written in terms of $\{\mu_1, \mu_2, \mu_3,\rho_T(1)\}$, the first three inter-event time moments  and the first-lag auto-correlation coefficient of the inter-event times. {The next} result establishes that, in order to completely characterize (\ref{K2}), two more moments involving the batch size, $\beta_1$ and $\eta$, as in (\ref{Eq:BMoments}) and (\ref{joint_moment}), respectively, should be added.

\begin{theorem}\label{TheoChaBMMPP2}
Let $\mathcal{B}_2 = \left\{\boldsymbol{D_0},\boldsymbol{D_1},\boldsymbol{D_2}\right\}$  be a representation of a $\bmmpp_2(2)$ as in (\ref{K2}). Then, $\mathcal{B}_2$ is completely characterized by the six moments $\{\mu_1,\mu_2,\mu_3,\rho_T(1),\beta_1,\eta\}$.
\end{theorem}

\begin{proof}
Let $\mathcal{M} = \left\{\boldsymbol{D_0},\boldsymbol{D_1}+\boldsymbol{D_2}\right\}$ be the $\mmpp_2$ associated to $\mathcal{B}_2 = \left\{\boldsymbol{D_0},\boldsymbol{D_1},\boldsymbol{D_2}\right\}$. From Lemma \ref{Le:FromCanFormToMMPP},  representation $\mathcal{M}$ can be rewritten as
\begin{eqnarray*}
\boldsymbol{D_0}&=&\left( \begin{array}{cc}
x(\mu_1,\mu_2,\mu_3,\rho_T) & y(\mu_1,\mu_2,\mu_3,\rho_T)  \\
r(\mu_1,\mu_2,\mu_3,\rho_T) & u(\mu_1,\mu_2,\mu_3,\rho_T)  \end{array} \right)\\
\boldsymbol{D_1}+\boldsymbol{D_2}&=&\left( \begin{array}{cc}
-x(\mu_1,\mu_2,\mu_3,\rho_T) -y(\mu_1,\mu_2,\mu_3,\rho_T) & 0  \\
0 & -r(\mu_1,\mu_2,\mu_3,\rho_T)-u(\mu_1,\mu_2,\mu_3,\rho_T)  \end{array} \right).
\end{eqnarray*}
Hence
\begin{eqnarray*}
\boldsymbol{D_0}&=&\left( \begin{array}{cc}
x(\mu_1,\mu_2,\mu_3,\rho_T) & y(\mu_1,\mu_2,\mu_3,\rho_T)  \\
r(\mu_1,\mu_2,\mu_3,\rho_T) & u(\mu_1,\mu_2,\mu_3,\rho_T)  \end{array} \right)\\
\boldsymbol{D_1}&=&\left( \begin{array}{cc}
w & 0  \\
0 & q  \end{array} \right)\\
\boldsymbol{D_2}&=&\left( \begin{array}{cc}
-x(\mu_1,\mu_2,\mu_3,\rho_T) -y(\mu_1,\mu_2,\mu_3,\rho_T) -w & 0  \\
0 & -r(\mu_1,\mu_2,\mu_3,\rho_T)-u(\mu_1,\mu_2,\mu_3,\rho_T) -q  \end{array} \right).
\end{eqnarray*}

The quantities $\beta_1$ and $\eta$ defined in (\ref{Eq:BMoments}) and (\ref{joint_moment}) respectively, can be written in the case of the $\bmmpp_2(K)$ as
{
\begin{eqnarray}\label{CharMMPPEqbeta1}
\beta_1&=&\frac{2(rx+\textcolor{black}{2ry}+yu) +rw+yq}{(rx+\textcolor{black}{2ry}+yu)}
\end{eqnarray}}
and 
\begin{eqnarray}\label{CharMMPPEqETB}
\eta&=&\frac{rw(y-u)+qy(r-x)+(ry-xu)(2r+2y)}{(rx+2ry+yu)(xu-ry)}.
\end{eqnarray}
From (\ref{CharMMPPEqbeta1}) 
\begin{equation}\label{CharBMMPP2eqrw}
rw= (\beta_1-2)(rx+ry+ry+yu)-yq
\end{equation}
and from substituting (\ref{CharBMMPP2eqrw}) in (\ref{CharMMPPEqETB}),
\begin{eqnarray*}
\eta&=&\frac{[(\beta_1-2)(rx+2ry+yu)-yq](y-u)+qy(r-x)+(ry-xu)(2r+2y)}{(rx+2ry+yu)(xu-ry)}\\
&=&\frac{(\beta_1-2)(rx+2ry+yu)(y-u)+qy(r+u-x-y)+(ry-xu)(2r+2y)}{(rx+2ry+yu)(xu-ry)}.
\end{eqnarray*}
Hence
\begin{eqnarray}\label{CharBMMPP2eqq}
q&=&\frac{\eta(rx+2ry+yu)(xu-ry)-(\beta_1-2)(rx+2ry+yu)(y-u)-(ry-xu)(2r+2y)}{y(r+u-x-y)}\nonumber\\
&=&\frac{(rx+2ry+yu)[(xu-ry)\eta-(y-u)(\beta_1-2)]-(ry-xu)(2r+2y)}{y(r+u-x-y)}
\end{eqnarray}
and from substituting (\ref{CharBMMPP2eqq}) in (\ref{CharBMMPP2eqrw}),  $w$ is finally found as
\begin{equation*}\label{CharBMMPP2eqw}
w=\frac{(rx+2ry+yu)[(\beta_1-2)(r-x)-(xu-ry)\eta]+(ry-xu)(2r+2y)}{r(r+u-x-y)}
\end{equation*}
Since the parameters defining $\mathcal{B}_2$ are written in terms of the moments $\{\mu_1,\mu_2,\mu_3,\rho_T(1),\beta_1,\eta\}$, the proof is completed.
\end{proof}

\subsection{The case $K\geq 3$}\label{K3}
In this section the characterization in terms of moments is extended from the case $K=2$ to the case with an arbitrary maximum batch size $K$. The key for such generalization is the fact that given a $\bmmpp_2(K)$ represented by $ \mathcal{B}_K = \left\{\boldsymbol{D_0},\boldsymbol{D_1},...,\boldsymbol{D_K}\right\}$, then  {$K$} different $\bmmpp_2(2)$s can be obtained as
\begin{equation}\label{sub_bmmpp2}
\mathcal{B}_2^{(i)}=\{\boldsymbol{D_0},\boldsymbol{D_i}, \sum_{k \neq i} \boldsymbol{D_k}\}, \quad \textcolor{black}{i=1,...,K}.
\end{equation}

\begin{theorem}\label{TheoChaBMMPPK}
Let $ \mathcal{B}_K = \left\{\boldsymbol{D_0},\boldsymbol{D_1},...,\boldsymbol{D_K}\right\}$  be {the} representation of a $\bmmpp_2(K)$. Then, $\mathcal{B}_K$ is characterized by the set of $(2K+2)$ moments
\begin{equation}\label{moments_char}
\left\{\mu_1,\mu_2,\mu_3,\rho_T(1),\beta_1^{(1)},\eta^{(1)},\ldots,\beta_1^{(K-1)},\eta^{(K-1)}\right\},\end{equation}
where $\beta_1^{(i)}$ and $\eta^{(i)}$ are the moments defined according to (\ref{Eq:BMoments}) and (\ref{joint_moment}) of  $\mathcal{B}_2^{(i)}$, for $i=1,\ldots,(K-1)$, that is the $\bmmpp_2(2)$ as in (\ref{sub_bmmpp2}).
\end{theorem}
\begin{proof}
The proof is straightforward by applying Theorem \ref{TheoChaBMMPP2} to each one of the $\bmmpp_2(2)$s defined by  $\mathcal{B}_2^{(i)}$, as in (\ref{sub_bmmpp2}), for $i=1,\ldots,(K-1)$.
\end{proof}

\section{Inference for the $\bmmpp_2(K)$}\label{fitting}
In this section, an approach for estimating the parameters of a $\bmmpp_2(K)$ given observed inter-event times, $\boldsymbol{t} = (t_1, t_2, . . . , t_n)$ and batch sizes,  $\boldsymbol{b} = (b_1, b_2, . . . , b_n)$, is proposed. \textcolor{black}{This implies that some components of the process as the complete sequence of transition times and the sequence of visited states in the underlying Markov process are not observed, which corresponds with what usually occurs in practice.}

\textcolor{black}{Section \ref{fitting2} presents in detail the novel fitting algorithm, where the rate matrices $\boldsymbol{D_0},\ldots, \boldsymbol{D_K}$ are sequentially estimated via $(K+1)$ optimization problems solved by standard optimization routines. Then, Section \ref{simu} illustrates the performance of the method on simulated data sets and Section \ref{EM} compares the novel approach with an EM-based strategy proposed in the literature. Finally, Section \ref{teletraffic} addresses the modeling of the well-known Bellcore Aug89 data set, where in addition, a performance analysis related to the $\bmmpp/M/1$ queueing system is considered.}

\subsection{The fitting algorithm}\label{fitting2}
Theorem \ref{TheoChaBMMPPK} shows that any $\bmmpp_2(K)$, with rate matrix representation as in (\ref{representation}), is characterized by the set of $2(K+1)$ moments given by  (\ref{moments_char}). Specifically, such moments are given by $\mu_1$, $\mu_2$, $\mu_3$, and $\rho_T(1)$ \textcolor{black}{(concerning the inter-event time distribution)}, $\beta_1^{(1)},\ldots,\beta_1^{(K-1)}$ \textcolor{black}{(related to the batch size distribution)}, and $\eta^{(1)},\ldots,\eta^{(K-1)}$ \textcolor{black}{(joint moments concerning times and sizes)}. \cite{Carpepa} derive a moments matching method for estimating the parameters of a $\map_2$ as in (\ref{map2}), given a sequence of inter-event times $\boldsymbol{t} = (t_1, t_2, . . . , t_n)$.
A modified version of the approach in \cite{Carpepa} shall constitute the first step in our sequential fitting algorithm aimed to estimate matrix $\boldsymbol{D_0}$. Any $\bmmpp_2(K)$ given by $\mathcal{B}_2=\{\boldsymbol{D_0},\boldsymbol{D_1},\ldots, \boldsymbol{D_K}\}$ defines a $\mmpp_2$ represented by $\mathcal{M}=\{\boldsymbol{G_0}=\boldsymbol{D_0},\ \boldsymbol{G_1}=\sum_{k=1}^K\boldsymbol{D_k}\}$, where $\boldsymbol{G_0}$ and $\boldsymbol{G_1}$ are as in (\ref{mmpp2}); therefore, $\boldsymbol{G_0}$ ($\boldsymbol{D_0}$) can be estimated by the solution of the following moments matching optimization problem ($P0$):
\begin{equation*}(P0)
\left\{
\begin{array}{lll}
\underset{x,y,r,u}{\min} & \displaystyle \delta_{0,\textcolor{black}{\tau}}\color{black}(x,y,r,u) \\
\mbox{s.t.} & x,u\leq 0, \\
 & y,r\geq 0,\\
 &x+y\leq 0,\\
 &r+u\leq 0,
\end{array}
\right.
\end{equation*}
where the objective function is 
\begin{eqnarray*}
\!\!\!\!\!\!\delta_{0,\textcolor{black}{\tau}}\color{black}(x,y,r,u) &=& \left\{\rho_T(1)-\bar{\rho}_T(1) \right\}^2 +\\
&+&
 \textcolor{black}{\tau} \color{black} \left\{\left(\frac{\mu_1-\bar{\mu}_1}{\bar{\mu}_1}\right)^2   + \left(\frac{\mu_2-\bar{\mu}_2}{\bar{\mu}_2}\right)^2 +\left(\frac{\mu_3-\bar{\mu}_3}{\bar{\mu}_3}\right)^2  \right\},
\end{eqnarray*}
for a value of \textcolor{black}{{$\tau$}} to be tuned in practice, and where $\bar{\mu}_i$, for $i=1,2,3$ and $\bar{\rho}_T(1)$  denote the empirical moments (computed from the sample $\boldsymbol{t}$). Note that in the previous objective function, $\rho_T(1)=\rho_T(1)(x,y,r,u)$, and $\mu_i=\mu_i(x,y,r,u)$, for $i=1,2,3$.

Once $\boldsymbol{\hat{D}_0}$ is obtained as the solution of (P0), then, in order to estimate $\boldsymbol{D_1}$ (or equivalently ${w}_1,{q}_1$), consider (\ref{sub_bmmpp2}) for $i=1$, that is, the $\bmmpp_2(2)$ represented by $\mathcal{B}_2^{(1)}=\{\boldsymbol{D_0},\ \boldsymbol{D_1},\ \boldsymbol{D_2}+\ldots+\boldsymbol{D_K}\}$ and the optimization problem
\begin{equation*}(P1)
\left\{
\begin{array}{lll}
\underset{w_1,q_1}{\min} & \displaystyle \delta_{1,\textcolor{black}{\tau}}\color{black}(\hat{x},\hat{y},\hat{u},\hat{v},w_1,q_1) \\
\mbox{s.t.} & 0\leq w_1\leq -\hat{x}-\hat{y}, \\
 &0\leq q_1\leq -\hat{r}-\hat{u},
\end{array}
\right.
\end{equation*}
where, according to (\ref{representation}), $\hat{x},\hat{y},\hat{r},\hat{u}$ are the elements of $\boldsymbol{\hat{D}_0}$ and
\begin{equation}\label{of_D1}
\delta_{1,\textcolor{black}{\tau}}\color{black}(x,y,u,v,w_1,q_1) = \textcolor{black}{\tau}\color{black}\left\{\left(\frac{\beta_1^{(1)}-\bar{\beta}_{1}^{(1)}}{\bar{\beta}_{1}^{(1)}}\right)^2   + \left(\frac{\eta^{(1)}-\bar{\eta}^{(1)}}{\bar{\eta}^{(1)}}\right)^2 \right\}.
\end{equation}
In the previous objective function (\ref{of_D1}), $\beta_1^{(1)}=\beta_1^{(1)}(x,y,u,r,w_1,q_1)$ and 
similarly, $\eta^{(1)}=\eta^{(1)}(x,y,u,r,w_1,q_1)$. It is crucial to remark that, in order to compute the empirical moments $\bar{\beta}_{1}^{(1)}$ and $\bar{\eta}^{(1)}$, all batch sizes in $\boldsymbol{b} $ larger than $2$ are considered as equal to $2$. Once $\hat{w}_1$ and $\hat{q}_1$ are obtained as the solutions of ($P1$), the approach will be repeated for estimating $\boldsymbol{D_2}$ (using the representation of $\mathcal{B}_2^{(2)}$), $\boldsymbol{D_3}$,..., and finally $\boldsymbol{D_K}$. The algorithm is summarized in Table \ref{tab:1}. \bigskip

\begin{table}[h!]
\begin{center}
{\tt 
\begin{enumerate}
\item[1.] Obtain $(\hat{x},\hat{y},\hat{r},\hat{u})$ (equivalently, $\boldsymbol{\hat{D}_0}$) as the solution of (P0).
\item[2.] For $k =1,\ldots,K-1$ repeat:
\begin{enumerate}
\item[(a)] Compute the empirical moments $\bar{\beta}_1^{(k)}$ and $\bar{\eta}^{(k)}$ from $\boldsymbol{t}$ and the sample of baches $\boldsymbol{b}^{\star}=(b^{\star}_1,\ldots,b^{\star}_n)$, where for $j=1,\ldots,n$, $b^{\star}_j=1$ if $b_j=k$, or $b^{\star}_j=2$, otherwise.
\item[(b)] From $\boldsymbol{\hat{D}_0},\ldots,\boldsymbol{\hat{D}_{k-1}}$ and the moments $\hat{\beta}_1^{(k)}$, $\hat{\eta}^{(k)}$, obtain $\hat{w}_k,\hat{q}_k$ ($\boldsymbol{\hat{D}_k}$) as the solutions of 
\begin{equation*}(Pk)
\left\{
\begin{array}{lll}
\underset{w_k,q_k}{\min} & \displaystyle \delta_{k,\textcolor{black}{\tau}}\color{black}(\hat{x},\hat{y},\hat{u},\hat{v},\hat{w}_1,\hat{q}_1,\ldots,\hat{w}_{k-1},\hat{q}_{k-1},w_k,q_k) \\
\mbox{s.t.} & 0\leq w_k\leq -\left(\hat{x}+\hat{y}+\hat{w}_1+\ldots+\hat{w}_{k-1}\right), \\
 &0\leq q_k\leq -\left(\hat{r}+\hat{u}+\hat{q}_1+\ldots+\hat{q}_{k-1}\right),
\end{array}
\right.
\end{equation*}
where 
$$
\delta_{k,\textcolor{black}{\tau}}\color{black}(w_k,q_k) = \textcolor{black}{\tau}\color{black}\left\{\left(\frac{\beta_1^{(k)}-\bar{\beta}_{1}^{(k)}}{\bar{\beta}_{1}^{(k)}}\right)^2   + \left(\frac{\eta^{(k)}-\bar{\eta}^{(k)}}{\bar{\eta}^{(k)}}\right)^2 \right\}.
$$
\end{enumerate}
\end{enumerate}
}
\end{center}
\caption{\label{tab:1} Sequential algorithm for estimating the $\bmmpp_2(K)$ parameters}
\end{table}

It is important to comment that the optimization problems ($Pk$) in Table \ref{tab:1}, for $k=0,\ldots,K-1$ are straightforward problems in two variables each, solved using standard optimization routines ( {\tt fmincon} in  MATLAB${}^\copyright$), where a multistart with $100$ randomly
chosen starting points was executed.

\textcolor{black}{\subsection{A simulational study}\label{simu}
The aim of this section is twofold: on one hand, the behavior of the sequential algorithm described in Section \ref{fitting} is illustrated on the base of two simulated data sets and, on the other hand, a sensitivity analysis concerning the tuning parameter $\tau$ is undertaken. Each simulated data set consists in a sequence of inter-event times $\boldsymbol{t} = (t_1, t_2, . . . , t_n)$ and a sequence of batch sizes $\boldsymbol{b} = (b_1, b_2, . . . , b_n)$. The first data set was simulated from the $\bmmpp_2(2)$ represented by the rate matrices $\{\boldsymbol{D_0},\boldsymbol{D_1}, \boldsymbol{D_2}\}$ shown in the second column of the top part of Table \ref{ta:comp_tau}; the second one was generated from the $\bmmpp_2(4)$ characterized by $\{\boldsymbol{D_0},\boldsymbol{D_1}, \boldsymbol{D_2},\boldsymbol{D_3},\boldsymbol{D_4}\}$ as in the second column of the bottom part of Table \ref{ta:comp_tau}. An important remark concerning the samples sizes $n$ needs to be made at this point. The estimation approach proposed in Section \ref{fitting} uses as input arguments a set of empirical moments concerning both the inter-event times and batch sizes. Since the process is known to be identifiable \citep{Yera} then, the closer the empirical moments are to the theoretical moments, the more accurate the estimated parameters will be. Therefore, the issue of the sample size is critical in this context. In this paper, we adopt the approach as in \cite{BayesPepa}, where the coefficient of variation of the inter-event times is taken into account. Specifically, if the coefficient of variation is high, then the sequence of inter-event times will present more variability and therefore, the approximation of the empirical moments to the theoretical ones may be poor. Under the two considered generator processes, the coefficients of variation are equal to $1.02$ and $2.048$, respectively.  Similarly as in \cite{BayesPepa}, we fix a lower value of the sample size ($n=300$) for the first case than for the second case ($n=1000$).}

\textcolor{black}{The results obtained when the novel estimation approach is used to fit the traces are shown in Table \ref{ta:comp_tau}, where the top part is related to the simulated sample from the $\bmmpp_2(2)$ while the bottom part concerns the second simulated sample from the $\bmmpp_2(4)$. The second column in the Table shows the generator process and the characterizing theoretical moments according to Sections 3.2 and 3.3. The third column in the table shows the empirical moments from the simulated traces. The rest of columns show the estimated rate matrices and estimated characterizing moments under the new approach, for an assortment of values of the tuning parameter $\tau$ ($\tau \in \{0.001,\ 0.01,\ 0.1,\ 1,\ 10,\ 100\}$). Finally, the last row shows the running time (measured in seconds) employed for the novel method in an Intel Core i5 of dual-core 2.6 GHz processor with 4Gb of memory ram (for a prototype code written in MATLAB${}^\copyright$). }

\textcolor{black}{Some comments arise from the results presented in Table \ref{ta:comp_tau}. First, from the third column it can be concluded that the selected samples sizes ($n=300$ and $n=1000$) are good enough to guarantee an accurate approximation of the empirical moments to the theoretical ones. Second, the value of $\tau$ does not seem to affect significantly the estimation: both the rate matrices and estimated moments are close to the real ones in all cases. However, the value of $\tau$ seems to have an impact on the computational time: the lower $\tau$ is, the faster the method turns out. For this reason, the smallest tested value ($\tau=0.001$) will be considered from now on in the rest of experiments. Finally, it is important to note that the sample size does not affect the running times, a fact which in any case was expected since the input arguments of the algorithm are empirical moments (and not the original traces).}

\textcolor{black}{The choice of processes in Table \ref{ta:comp_tau}  is also related to the Remark \ref{rosa}. As can be observed, the first process presents autocorrelation between batches, autocorrelation between the inter-arrivals times and correlation between T and B close to zero; while in the second process, they are significantly different from zero. The difference between fitting a BMMPP as defined in this work and the MAP with i.i.d. batch arrivals  is  more relevant in the second model than in the first as is illustrated in Table \ref{ta:comp_bmmppvsiid}. It can be seen that the quality of the performance in the adjustment is better using the methodology developed in this paper while the computational times are very similar.}

\begingroup 
\begin{table}[!h]
\sisetup{ 
	detect-all,
	table-number-alignment = center,
	table-figures-integer = 1,
	table-figures-decimal = 3,
	explicit-sign
} 
\resizebox{1.00\textwidth}{!}{
	\begin{tabular}{|c|c|c|c|c|c|c|c|c|}
	\hline
	& &&\multicolumn{6}{c|}{$\tau$}\\
	\hline
	&  \multicolumn{1}{c|}{$\begin{array}{c} Generator\\ Process\end{array}$}&  \multicolumn{1}{c|}{\emph{Empirical}}&0.001&0.01 &   0.1 &  1 & 10 & 100    \\ 
	\hline
	&&&&&&&&\\
	$\boldsymbol{D_0}$&$\left( \begin{array}{cc} \tablenum[table-format=2]{-5} & \tablenum[table-format=3]{2}  \\ \tablenum[table-format=2]{5} & \tablenum[table-format=3]{10} \end{array}\right)$&
	-&
	$\left( \begin{array}{cc}\tablenum[table-format=2.2]{-5.95} &
	\tablenum[table-format=3.2]{3.07}\\ \tablenum[table-format=2.2]{6.59} & \tablenum[table-format=3.2]{-11.64} \end{array}\right)$&
	$\left( \begin{array}{cc}\tablenum[table-format=2.2]{-5.95} &
	\tablenum[table-format=3.2]{3.08}\\  \tablenum[table-format=2.2]{6.46} & \tablenum[table-format=3.2]{-11.50} \end{array}\right)$&  
	$\left( \begin{array}{cc}\tablenum[table-format=2.2]{-5.92} &
	\tablenum[table-format=3.2]{3.04}\\  \tablenum[table-format=2.2]{6.52} & \tablenum[table-format=3.2]{-11.57} \end{array}\right)$&
	$\left( \begin{array}{cc} \tablenum[table-format=2.2]{-6.22} & \tablenum[table-format=3.2]{3.40}\\
	\tablenum[table-format=2.2]{6.30} & \tablenum[table-format=3.2]{-11.25} \end{array}\right)$&
	$\left( \begin{array}{cc} \tablenum[table-format=2.2]{-5.86} & \tablenum[table-format=3.2]{2.98}\\
	\tablenum[table-format=2.2]{6.60} & \tablenum[table-format=3.2]{-11.68} \end{array}\right)$&
	$\left( \begin{array}{cc} \tablenum[table-format=2.2]{-5.78} & \tablenum[table-format=3.2]{2.87}\\
	\tablenum[table-format=2.2]{7.64} & \tablenum[table-format=3.2]{-12.95} \end{array}\right)$\\
	&&&&&&&&\\
	$\boldsymbol{D_1}$&
	$\left( \begin{array}{cc}\tablenum[table-format=2]{1}&
	\tablenum[table-format=3]{0}\\
	\tablenum[table-format=2]{0}&
	\tablenum[table-format=3]{2}\end{array}\right)$&
	-&
	$\left( \begin{array}{cc}\tablenum[table-format=2.2]{0.91}&
	\tablenum[table-format=3.2]{0}\\
	\tablenum[table-format=2.2]{0}&
	\tablenum[table-format=3.2]{2.08}\end{array}\right)$&
	$\left( \begin{array}{cc}\tablenum[table-format=2.2]{0.91} &
	\tablenum[table-format=3.2]{0}\\ 
	\tablenum[table-format=2.2]{0} &
	\tablenum[table-format=3.2]{2.07}\end{array}\right)$& 
	$\left(\begin{array}{cc}\tablenum[table-format=2.2]{0.91} & \tablenum[table-format=3.2]{0} \\
	\tablenum[table-format=2.2]{0} & \tablenum[table-format=3.2]{2.08}\end{array}\right)$&
	$\left(\begin{array}{cc}\tablenum[table-format=2.2]{0.89} & \tablenum[table-format=3.2]{0} \\
	\tablenum[table-format=2.2]{0} & \tablenum[table-format=3.2]{2.02}\end{array}\right)$&
	$\left(\begin{array}{cc}\tablenum[table-format=2.2]{0.92} & \tablenum[table-format=3.2]{0}\\
	\tablenum[table-format=2.2]{0} & \tablenum[table-format=3.2]{2.09}\end{array}\right)$&
	$\left(\begin{array}{cc}\tablenum[table-format=2.2]{0.93} & \tablenum[table-format=3.2]{0}\\
	\tablenum[table-format=2.2]{0} & \tablenum[table-format=3.2]{2.23}\end{array}\right)$\\
	&&&&&&&&\\
	$\boldsymbol{D_2}$&
	$\left( \begin{array}{cc}\tablenum[table-format=2]{2}&
	\tablenum[table-format=3]{0}\\
	\tablenum[table-format=2]{0}&
	\tablenum[table-format=3]{3}\end{array}\right)$&
	-&
	$\left( \begin{array}{cc}\tablenum[table-format=2.2]{1.96} &
	\tablenum[table-format=2.2]{0} \\
	\tablenum[table-format=3.2]{0} &
	\tablenum[table-format=2.2]{2.98}\end{array}\right)$&
	$\left( \begin{array}{cc} \tablenum[table-format=2.2]{1.96} & \tablenum[table-format=3.2]{0}\\
	\tablenum[table-format=2.2]{0} & \tablenum[table-format=3.2]{2.96}\end{array}\right)$&
	$\left( \begin{array}{cc} \tablenum[table-format=2.2]{1.96} & \tablenum[table-format=3]{0}\\
	\tablenum[table-format=2.2]{0} & \tablenum[table-format=3.2]{2.97}\end{array}\right)$&
	$\left( \begin{array}{cc}\tablenum[table-format=2.2]{1.94} & \tablenum[table-format=3.2]{0}\\
	\tablenum[table-format=2.2]{0} & \tablenum[table-format=3.2]{2.92}\end{array}\right)$&
	$\left(\begin{array}{cc}\tablenum[table-format=2.2]{1.97}& \tablenum[table-format=3.2]{0}\\
	\tablenum[table-format=2.2]{0} & \tablenum[table-format=3.2]{2.99}\end{array}\right)$&
	$\left(\begin{array}{cc}\tablenum[table-format=2.2]{1.98}& \tablenum[table-format=3.2]{0}\\
	\tablenum[table-format=2.2]{0} & \tablenum[table-format=3.2]{3.08}\end{array}\right)$\\
	&&&&&&&&\\
	$\mu_1$&\multicolumn{1}{c|}{0.28}&
	\multicolumn{1}{c|}{0.28}&
	\multicolumn{1}{c|}{0.28}&
	\multicolumn{1}{c|}{0.28}&
	\multicolumn{1}{c|}{0.28}&
	\multicolumn{1}{c|}{0.28}&
	\multicolumn{1}{c|}{0.28}&
	\multicolumn{1}{c|}{0.28}\\
	$\mu_2$&\multicolumn{1}{c|}{0.16}&
	\multicolumn{1}{c|}{0.16}&
	\multicolumn{1}{c|}{0.16}&
	\multicolumn{1}{c|}{0.16}&
	\multicolumn{1}{c|}{0.16}&
	\multicolumn{1}{c|}{0.16}&
	\multicolumn{1}{c|}{0.16}&
	\multicolumn{1}{c|}{0.16}\\
	$\mu_3$&\multicolumn{1}{c|}{0.138}&
	\multicolumn{1}{c|}{0.138}&
	\multicolumn{1}{c|}{0.138}&
	\multicolumn{1}{c|}{0.138}&
	\multicolumn{1}{c|}{0.138}&
	\multicolumn{1}{c|}{0.138}&
	\multicolumn{1}{c|}{0.138}&
	\multicolumn{1}{c|}{0.138}\\
	$\rho_T(1)$&\multicolumn{1}{c|}{\tablenum[table-format=1.2e2]{7.35e-3}}&
	\multicolumn{1}{c|}{\tablenum[table-format=1.2e2]{5.99e-3}}&
	\multicolumn{1}{c|}{\tablenum[table-format=1.2e2]{5.99e-3}}&
	\multicolumn{1}{c|}{\tablenum[table-format=1.2e2]{6.01e-3}}&
	\multicolumn{1}{c|}{\tablenum[table-format=1.2e2]{6.03e-3}}&
	\multicolumn{1}{c|}{\tablenum[table-format=1.2e2]{5.81e-3}}&
	\multicolumn{1}{c|}{\tablenum[table-format=1.2e2]{6.07e-3}}&
	\multicolumn{1}{c|}{\tablenum[table-format=1.2e2]{5.99e-3}}\\
	$\beta_1$&\multicolumn{1}{c|}{1.64}&
	\multicolumn{1}{c|}{1.64}&
	\multicolumn{1}{c|}{1.64}&
	\multicolumn{1}{c|}{1.64}&
	\multicolumn{1}{c|}{1.64}&
	\multicolumn{1}{c|}{1.64}&
	\multicolumn{1}{c|}{1.64}&
	\multicolumn{1}{c|}{1.64}\\
	$\eta_1$&\multicolumn{1}{c|}{0.46}&
	\multicolumn{1}{c|}{0.46}&
	\multicolumn{1}{c|}{0.46}&
	\multicolumn{1}{c|}{0.46}&
	\multicolumn{1}{c|}{0.46}&
	\multicolumn{1}{c|}{0.46}&
	\multicolumn{1}{c|}{0.46}&
	\multicolumn{1}{c|}{0.46}\\
	$\rho_B(1)$&\multicolumn{1}{c|}{\tablenum[table-format=1.2e3]{2.24e-3}}&
\multicolumn{1}{c|}{\tablenum[table-format=1.2e3]{1.73e-3}}&
\multicolumn{1}{c|}{\tablenum[table-format=1.2e3]{1.02e-3}}&
\multicolumn{1}{c|}{\tablenum[table-format=1.2e3]{8.82e-4}}&
\multicolumn{1}{c|}{\tablenum[table-format=1.2e3]{8.28e-4}}&
\multicolumn{1}{c|}{\tablenum[table-format=1.2e3]{8.30e-4}}&
\multicolumn{1}{c|}{\tablenum[table-format=1.2e3]{8.39e-4}}&
\multicolumn{1}{c|}{\tablenum[table-format=1.2e3]{8.39e-4}}\\
	$corr(T,B)$&\multicolumn{1}{c|}{\tablenum[table-format=1.2e3]{5.83e-3}}&
	\multicolumn{1}{c|}{\tablenum[table-format=1.2e3]{7.53e-3}}&
	\multicolumn{1}{c|}{\tablenum[table-format=1.2e3]{7.02e-3}}&
	\multicolumn{1}{c|}{\tablenum[table-format=1.2e3]{7.51e-4}}&
	\multicolumn{1}{c|}{\tablenum[table-format=1.2e3]{7.50e-3}}&
	\multicolumn{1}{c|}{\tablenum[table-format=1.2e3]{7.53e-3}}&
	\multicolumn{1}{c|}{\tablenum[table-format=1.2e3]{7.53e-3}}&
	\multicolumn{1}{c|}{\tablenum[table-format=1.2e3]{7.53e-3}}\\
	&&&&&&&&\\
	$\begin{array}{c} running\\ time\end{array}$
	&\multicolumn{1}{c|}{-}&\multicolumn{1}{c|}{-}&\multicolumn{1}{c|}{21.59}&
	\multicolumn{1}{c|}{35.24}&
	\multicolumn{1}{c|}{51.40}&
	\multicolumn{1}{c|}{108.64}&
	\multicolumn{1}{c|}{180.69}&
	\multicolumn{1}{c|}{183.96}\\
	\hline
	\hline
	&&&&&&&&\\
	$\boldsymbol{D_0}$&$\left( \begin{array}{cc} \tablenum[table-format=2.2]{-0.58} & \tablenum[table-format=3.2]{0.09}  \\ \tablenum[table-format=2.2]{1.91} & \tablenum[table-format=3.2]{-14.20} \end{array}\right)$&
	-&
	$\left( \begin{array}{cc}\tablenum[table-format=2.2]{-0.58} &
	\tablenum[table-format=3.2]{0.09}\\ \tablenum[table-format=2.2]{1.85} & \tablenum[table-format=3.2]{-13.86} \end{array}\right)$&
	$\left( \begin{array}{cc}\tablenum[table-format=2.2]{-0.58} &
	\tablenum[table-format=3.2]{0.09}\\ \tablenum[table-format=2.2]{1.84} & \tablenum[table-format=3.2]{-13.82} \end{array}\right)$&  
	$\left( \begin{array}{cc}\tablenum[table-format=2.2]{-0.58} &
	\tablenum[table-format=3.2]{0.09}\\ \tablenum[table-format=2.2]{1.84} & \tablenum[table-format=3.2]{-13.79} \end{array}\right)$&
	$\left( \begin{array}{cc}\tablenum[table-format=2.2]{-0.58} &
	\tablenum[table-format=3.2]{0.09}\\ \tablenum[table-format=2.2]{1.84} & \tablenum[table-format=3.2]{-13.78} \end{array}\right)$&
	$\left( \begin{array}{cc}\tablenum[table-format=2.2]{-0.58} &
	\tablenum[table-format=3.2]{0.09}\\ \tablenum[table-format=2.2]{1.84} & \tablenum[table-format=3.2]{-13.79} \end{array}\right)$&
	$\left( \begin{array}{cc}\tablenum[table-format=2.2]{-0.58} &
	\tablenum[table-format=3.2]{0.09}\\ \tablenum[table-format=2.2]{1.83} & \tablenum[table-format=3.2]{-13.78} \end{array}\right)$\\
	&&&&&&&&\\
	$\boldsymbol{D_1}$&
	$\left( \begin{array}{cc}\tablenum[table-format=2.2]{0.08}&
	\tablenum[table-format=3.2]{0}\\
	\tablenum[table-format=2.2]{0}&
	\tablenum[table-format=3.2]{11.47}\end{array}\right)$&
	-&
	$\left( \begin{array}{cc}\tablenum[table-format=2.2]{0.08}&
	\tablenum[table-format=3.2]{0}\\
	\tablenum[table-format=2.2]{0}&
	\tablenum[table-format=3.2]{11.17}\end{array}\right)$&
	$\left( \begin{array}{cc}\tablenum[table-format=2.2]{0.08}&
	\tablenum[table-format=3.2]{0}\\
	\tablenum[table-format=2.2]{0}&
	\tablenum[table-format=3.2]{11.14}\end{array}\right)$& 
	$\left( \begin{array}{cc}\tablenum[table-format=2.2]{0.08}&
	\tablenum[table-format=3.2]{0}\\
	\tablenum[table-format=2.2]{0}&
	\tablenum[table-format=3.2]{11.14}\end{array}\right)$&
	$\left( \begin{array}{cc}\tablenum[table-format=2.2]{0.08}&
	\tablenum[table-format=3.2]{0}\\
	\tablenum[table-format=2.2]{0}&
	\tablenum[table-format=3.2]{11.14}\end{array}\right)$&
	$\left( \begin{array}{cc}\tablenum[table-format=2.2]{0.08}&
	\tablenum[table-format=3.2]{0}\\
	\tablenum[table-format=2.2]{0}&
	\tablenum[table-format=3.2]{11.14}\end{array}\right)$&
	$\left( \begin{array}{cc}\tablenum[table-format=1.2]{0.08}&
	\tablenum[table-format=3.2]{0}\\
	\tablenum[table-format=2.2]{0}&
	\tablenum[table-format=3.2]{11.14}\end{array}\right)$\\
	&&&&&&&&\\
	$\boldsymbol{D_2}$&
	$\left( \begin{array}{cc}\tablenum[table-format=1.2]{0.15}&
	\tablenum[table-format=3.1]{0}\\
	\tablenum[table-format=2.1]{0}&
	\tablenum[table-format=3.1]{0.10}\end{array}\right)$&
	-&
	$\left( \begin{array}{cc}\tablenum[table-format=2.2]{0.15} &
	\tablenum[table-format=2.2]{0} \\
	\tablenum[table-format=2.2]{0} &
	\tablenum[table-format=2.2]{0.08}\end{array}\right)$&
	$\left( \begin{array}{cc}\tablenum[table-format=2.2]{0.15} &
	\tablenum[table-format=2.2]{0} \\
	\tablenum[table-format=2.2]{0} &
	\tablenum[table-format=2.2]{0.07}\end{array}\right)$&
	$\left( \begin{array}{cc}\tablenum[table-format=2.2]{0.15} &
	\tablenum[table-format=2.2]{0} \\
	\tablenum[table-format=2.2]{0} &
	\tablenum[table-format=2.2]{0.07}\end{array}\right)$&
	$\left( \begin{array}{cc}\tablenum[table-format=2.2]{0.15} &
	\tablenum[table-format=2.2]{0} \\
	\tablenum[table-format=2.2]{0} &
	\tablenum[table-format=2.2]{0.07}\end{array}\right)$&
	$\left( \begin{array}{cc}\tablenum[table-format=2.2]{0.15} &
	\tablenum[table-format=2.2]{0} \\
	\tablenum[table-format=2.2]{0} &
	\tablenum[table-format=2.2]{0.07}\end{array}\right)$&
	$\left( \begin{array}{cc}\tablenum[table-format=2.2]{0.15} &
	\tablenum[table-format=2.2]{0} \\
	\tablenum[table-format=2.2]{0} &
	\tablenum[table-format=2.2]{0.07}\end{array}\right)$\\
	&&&&&&&&\\
	$\boldsymbol{D_3}$&
	$\left( \begin{array}{cc}\tablenum[table-format=2.1]{0.25}&
	\tablenum[table-format=3.1]{0}\\
	\tablenum[table-format=2.1]{0}&
	\tablenum[table-format=3.1]{0.60}\end{array}\right)$&
	-&
	$\left( \begin{array}{cc}\tablenum[table-format=2.2]{0.25} &
	\tablenum[table-format=2.2]{0} \\
	\tablenum[table-format=2.2]{0} &
	\tablenum[table-format=2.2]{0.64}\end{array}\right)$&
	$\left( \begin{array}{cc}\tablenum[table-format=2.2]{0.25} &
	\tablenum[table-format=2.2]{0} \\
	\tablenum[table-format=2.2]{0} &
	\tablenum[table-format=2.2]{0.64}\end{array}\right)$&
	$\left( \begin{array}{cc}\tablenum[table-format=2.2]{0.25} &
	\tablenum[table-format=2.2]{0} \\
	\tablenum[table-format=2.2]{0} &
	\tablenum[table-format=2.2]{0.64}\end{array}\right)$&
	$\left( \begin{array}{cc}\tablenum[table-format=2.2]{0.25} &
	\tablenum[table-format=2.2]{0} \\
	\tablenum[table-format=2.2]{0} &
	\tablenum[table-format=2.2]{0.64}\end{array}\right)$&
	$\left( \begin{array}{cc}\tablenum[table-format=2.2]{0.25} &
	\tablenum[table-format=2.2]{0} \\
	\tablenum[table-format=2.2]{0} &
	\tablenum[table-format=2.2]{0.64}\end{array}\right)$&
	$\left( \begin{array}{cc}\tablenum[table-format=2.2]{0.25} &
	\tablenum[table-format=2.2]{0} \\
	\tablenum[table-format=2.2]{0} &
	\tablenum[table-format=2.2]{0.64}\end{array}\right)$\\	
	&&&&&&&&\\
	$\boldsymbol{D_4}$&
	$\left( \begin{array}{cc}\tablenum[table-format=2.1]{0.01}&
	\tablenum[table-format=3.1]{0}\\
	\tablenum[table-format=2.1]{0}&
	\tablenum[table-format=3.1]{0.12}\end{array}\right)$&
	-&
	$\left( \begin{array}{cc}\tablenum[table-format=2.2]{0.01} &
	\tablenum[table-format=2.2]{0} \\
	\tablenum[table-format=2.2]{0} &
	\tablenum[table-format=2.2]{0.10}\end{array}\right)$&
	$\left( \begin{array}{cc}\tablenum[table-format=2.2]{0.01} &
	\tablenum[table-format=2.2]{0} \\
	\tablenum[table-format=2.2]{0} &
	\tablenum[table-format=2.2]{0.11}\end{array}\right)$&
	$\left( \begin{array}{cc}\tablenum[table-format=2.2]{0.01} &
	\tablenum[table-format=2.2]{0} \\
	\tablenum[table-format=2.2]{0} &
	\tablenum[table-format=2.2]{0.11}\end{array}\right)$&
	$\left( \begin{array}{cc}\tablenum[table-format=2.2]{0.01} &
	\tablenum[table-format=2.2]{0} \\
	\tablenum[table-format=2.2]{0} &
	\tablenum[table-format=2.2]{0.11}\end{array}\right)$&
	$\left( \begin{array}{cc}\tablenum[table-format=2.2]{0.01} &
	\tablenum[table-format=2.2]{0} \\
	\tablenum[table-format=2.2]{0} &
	\tablenum[table-format=2.2]{0.11}\end{array}\right)$&
	$\left( \begin{array}{cc}\tablenum[table-format=2.2]{0.01} &
	\tablenum[table-format=2.2]{0} \\
	\tablenum[table-format=2.2]{0} &
	\tablenum[table-format=2.2]{0.11}\end{array}\right)$\\
	&&&&&&&&\\
	$\mu_1$&\multicolumn{1}{c|}{0.98}&
	\multicolumn{1}{c|}{0.97}&
	\multicolumn{1}{c|}{0.97}&
	\multicolumn{1}{c|}{0.97}&
	\multicolumn{1}{c|}{0.97}&
	\multicolumn{1}{c|}{0.97}&
	\multicolumn{1}{c|}{0.97}&
	\multicolumn{1}{c|}{0.97}\\
	$\mu_2$&\multicolumn{1}{c|}{3.34}&
	\multicolumn{1}{c|}{3.26}&
	\multicolumn{1}{c|}{3.26}&
	\multicolumn{1}{c|}{3.26}&
	\multicolumn{1}{c|}{3.26}&
	\multicolumn{1}{c|}{3.26}&
	\multicolumn{1}{c|}{3.26}&
	\multicolumn{1}{c|}{3.26}\\
	$\mu_3$&\multicolumn{1}{c|}{17.65}&
	\multicolumn{1}{c|}{17.08}&
	\multicolumn{1}{c|}{17.08}&
	\multicolumn{1}{c|}{17.08}&
	\multicolumn{1}{c|}{17.08}&
	\multicolumn{1}{c|}{17.08}&
	\multicolumn{1}{c|}{17.08}&
	\multicolumn{1}{c|}{17.08}\\
	$\rho_T(1)$&\multicolumn{1}{c|}{0.22}&
	\multicolumn{1}{c|}{0.22}&
	\multicolumn{1}{c|}{0.22}&
	\multicolumn{1}{c|}{0.22}&
	\multicolumn{1}{c|}{0.22}&
	\multicolumn{1}{c|}{0.22}&
	\multicolumn{1}{c|}{0.22}&
	\multicolumn{1}{c|}{0.22}\\
	$\beta_1$&\multicolumn{1}{c|}{1.42}&
	\multicolumn{1}{c|}{1.42}&
	\multicolumn{1}{c|}{1.42}&
	\multicolumn{1}{c|}{1.42}&
	\multicolumn{1}{c|}{1.42}&
	\multicolumn{1}{c|}{1.42}&
	\multicolumn{1}{c|}{1.42}&
	\multicolumn{1}{c|}{1.42}\\
	$\beta_2$&\multicolumn{1}{c|}{1.86}&
	\multicolumn{1}{c|}{1.86}&
	\multicolumn{1}{c|}{1.86}&
	\multicolumn{1}{c|}{1.86}&
	\multicolumn{1}{c|}{1.86}&
	\multicolumn{1}{c|}{1.86}&
	\multicolumn{1}{c|}{1.86}&
	\multicolumn{1}{c|}{1.86}\\
	$\beta_3$&\multicolumn{1}{c|}{1.73}&
	\multicolumn{1}{c|}{1.74}&
	\multicolumn{1}{c|}{1.74}&
	\multicolumn{1}{c|}{1.74}&
	\multicolumn{1}{c|}{1.74}&
	\multicolumn{1}{c|}{1.74}&
	\multicolumn{1}{c|}{1.74}&
	\multicolumn{1}{c|}{1.74}\\	$\eta_1$&\multicolumn{1}{c|}{1.67}&
	\multicolumn{1}{c|}{1.65}&
	\multicolumn{1}{c|}{1.65}&
	\multicolumn{1}{c|}{1.65}&
	\multicolumn{1}{c|}{1.65}&
	\multicolumn{1}{c|}{1.65}&
	\multicolumn{1}{c|}{1.65}&
	\multicolumn{1}{c|}{1.65}\\
	$\eta_1$&\multicolumn{1}{c|}{1.71}&
	\multicolumn{1}{c|}{1.68}&
	\multicolumn{1}{c|}{1.68}&
	\multicolumn{1}{c|}{1.68}&
	\multicolumn{1}{c|}{1.68}&
	\multicolumn{1}{c|}{1.68}&
	\multicolumn{1}{c|}{1.68}&
	\multicolumn{1}{c|}{1.68}\\
	$\eta_3$&\multicolumn{1}{c|}{1.53}&
	\multicolumn{1}{c|}{1.54}&
	\multicolumn{1}{c|}{1.52}&
	\multicolumn{1}{c|}{1.52}&
	\multicolumn{1}{c|}{1.52}&
	\multicolumn{1}{c|}{1.52}&
	\multicolumn{1}{c|}{1.52}&
	\multicolumn{1}{c|}{1.52}\\
	$\rho_B(1)$&\multicolumn{1}{c|}{0.36}&
	\multicolumn{1}{c|}{0.36}&
	\multicolumn{1}{c|}{0.28}&
	\multicolumn{1}{c|}{0.27}&
	\multicolumn{1}{c|}{0.27}&
	\multicolumn{1}{c|}{0.27}&
	\multicolumn{1}{c|}{0.27}&
	\multicolumn{1}{c|}{0.27}\\
	$Corr(T,B)$&\multicolumn{1}{c|}{0.33}&
	\multicolumn{1}{c|}{0.33}&
	\multicolumn{1}{c|}{0.33}&
	\multicolumn{1}{c|}{0.33}&
	\multicolumn{1}{c|}{0.33}&
	\multicolumn{1}{c|}{0.33}&
	\multicolumn{1}{c|}{0.33}&
	\multicolumn{1}{c|}{0.33}\\
	&&&&&&&&\\
	$\begin{array}{c} running\\ time\end{array}$
	&\multicolumn{1}{c|}{-}&\multicolumn{1}{c|}{-}&\multicolumn{1}{c|}{37.04}&
	\multicolumn{1}{c|}{65.88}&
	\multicolumn{1}{c|}{76.57}&
	\multicolumn{1}{c|}{159.58}&
	\multicolumn{1}{c|}{171.36}&
	\multicolumn{1}{c|}{179.02}\\
	\hline
	\hline
	\end{tabular}}
\vspace{0.05cm} 
\centering 
\caption{Performance of the novel sequential estimation method for a simulated trace from a $\bmmpp_2(2)$ (top part) and $\bmmpp_2(4)$ (bottom part) for an assortment of $\tau$ values.} \label{ta:comp_tau}
{\footnotesize
	\parbox{6.2in}{
		\medskip
		\begin{center}
		\end{center}}}
\end{table}
\endgroup

\begingroup 
\begin{table}[!h]
\sisetup{ 
	detect-all,
	table-number-alignment = center,
	table-figures-integer = 1,
	table-figures-decimal = 3,
	explicit-sign
} 
\resizebox{0.95\textwidth}{!}{
	\begin{tabular}{r|ccc|ccc}
	\hline
	& \multirow{ 2}{*}{Emp} & \multirow{ 2}{*}{Est BMMPP} & Est MMPP with  & \multirow{ 2}{*}{Emp} & \multirow{ 2}{*}{Est BMMPP} & Est MMPP with \\ 
	&  & & i.i.d. batches & & & i.i.d. batches  \\
	\hline
	 $\mu_1$& \tablenum[table-format=2.4]{0.2801} & \tablenum[table-format=2.4]{0.2797} & \tablenum[table-format=2.4]{0.2798}& \tablenum[table-format=2.4]{0.9657} & \tablenum[table-format=2.4]{0.9657} & \tablenum[table-format=2.4]{0.9656}\\
	$\mu_2$& \tablenum[table-format=2.4]{0.1602} & \tablenum[table-format=2.4]{0.1602} & \tablenum[table-format=2.4]{0.1604}& \tablenum[table-format=2.4]{3.2627} & \tablenum[table-format=2.4]{3.2628} & \tablenum[table-format=2.4]{3.2635}\\
	$\mu_3$& \tablenum[table-format=2.4]{0.1382} & \tablenum[table-format=2.4]{0.1382} & \tablenum[table-format=2.4]{1.381}& \tablenum[table-format=2.4]{17.0847} & \tablenum[table-format=2.4]{17.0842} & \tablenum[table-format=2.4]{17.0827}\\
	$\rho_T(1)$& \tablenum[table-format=2.3e2]{5.990e-3} & \tablenum[table-format=2.3e2]{5.989e-3}& \tablenum[table-format=2.3e2]{5.989e-3}& \tablenum[table-format=2.4]{0.2241} & \tablenum[table-format=2.4]{0.2241}& \tablenum[table-format=2.4]{0.2241}\\
	$\beta_1^{(1)}$& \tablenum[table-format=2.4]{1.6397} & \tablenum[table-format=2.4]{1.6397}& \tablenum[table-format=2.4]{1.6423}& \tablenum[table-format=2.4]{1.4182} & \tablenum[table-format=2.4]{1.4182}& \tablenum[table-format=2.4]{1.5358}\\
	$\eta^{(1)}$& \tablenum[table-format=2.4]{0.4603} & \tablenum[table-format=2.4]{0.4603}& \tablenum[table-format=2.4]{0.4595}& \tablenum[table-format=2.4]{1.6472} & \tablenum[table-format=2.4]{1.6472}& \tablenum[table-format=2.4]{1.4830}\\
	$\beta_1^{(2)}$&-&-&-& \tablenum[table-format=2.4]{1.8563} & \tablenum[table-format=2.4]{1.8558}& \tablenum[table-format=2.4]{1.7962}\\
	$\eta^{(2)}$&-&-&-& \tablenum[table-format=2.4]{1.6833} & \tablenum[table-format=2.4]{1.6835} & \tablenum[table-format=2.4]{1.7345}\\
	$\beta_1^{(3)}$&-&-&-& \tablenum[table-format=2.4]{1.7389} & \tablenum[table-format=2.4]{1.7390}
	& \tablenum[table-format=2.4]{1.6681}\\
	$\eta^{(3)}$&-&-&-& \tablenum[table-format=2.4]{1.5148} & \tablenum[table-format=2.4]{1.5148} & \tablenum[table-format=2.4]{1.6107}\\
	\hline 
	 $CV$& \tablenum[table-format=2.4]{1.0208} & \tablenum[table-format=2.4]{1.0218} & \tablenum[table-format=2.4]{1.0242}& \tablenum[table-format=2.4]{1.5806} & \tablenum[table-format=2.4]{1.5807} & \tablenum[table-format=2.4]{1.5812}\\
	$Skewness$& \tablenum[table-format=2.4]{5.9111} & \tablenum[table-format=2.4]{5.9004} & \tablenum[table-format=2.4]{5.8675}& \tablenum[table-format=2.4]{4.8034} & \tablenum[table-format=2.4]{4.8030}& \tablenum[table-format=2.4]{4.7997}\\
	 $Kurtosis$& \tablenum[table-format=2.4]{23.7610} & \tablenum[table-format=2.4]{27.7867} & \tablenum[table-format=2.4]{23.7164}& \tablenum[table-format=2.4]{22.0435} & \tablenum[table-format=2.4]{21.9988}
	& \tablenum[table-format=2.4]{21.9887}\\
	\hline 
	$\beta_1$& \tablenum[table-format=2.4]{1.6397} & \tablenum[table-format=2.4]{1.6397} & \tablenum[table-format=2.4]{1.6423}& \tablenum[table-format=2.4]{1.7064} & \tablenum[table-format=2.4]{1.7062}& \tablenum[table-format=2.4]{1.8679}\\
	 $\beta_2$& \tablenum[table-format=2.4]{2.9190} & \tablenum[table-format=2.4]{2.9190} & \tablenum[table-format=2.4]{2.9270}& \tablenum[table-format=2.4]{3.7224} & \tablenum[table-format=2.4]{3.7217}  & \tablenum[table-format=2.4]{4.2679}\\
	\hline
	$Corr(T,B)$& \tablenum[table-format=2.3e2]{7.536e-3} & \tablenum[table-format=2.3e2]{8.734e-3} & \tablenum[table-format=2.4]{0}& \tablenum[table-format=2.4]{0.3268} & \tablenum[table-format=2.4]{0.3270} & \tablenum[table-format=2.4]{0}\\
	$\rho_B(1)$& \tablenum[table-format=2.3e2]{2.248e-3} & \tablenum[table-format=2.3e2]{1.025e-3} & \tablenum[table-format=2.4]{0}& \tablenum[table-format=2.4]{0.3633} & \tablenum[table-format=2.4]{0.2665}& \tablenum[table-format=2.4]{0}\\
	$\rho_B(2)$& \tablenum[table-format=2.3e2]{3.577e-4} & \tablenum[table-format=2.3e2]{2.907e-4} & \tablenum[table-format=2.4]{0}& \tablenum[table-format=2.4]{0.2635} & \tablenum[table-format=2.4]{0.1991}& \tablenum[table-format=2.4]{0}\\
	$\rho_B(3)$& \tablenum[table-format=2.3e2]{7.479e-4} & \tablenum[table-format=2.3e2]{8.247e-5} & \tablenum[table-format=2.4]{0}& \tablenum[table-format=2.4]{0.2060} & \tablenum[table-format=2.4]{0.1488}& \tablenum[table-format=2.4]{0}\\
	$\rho_T(2)$& \tablenum[table-format=2.3e2]{6.426e-4} & \tablenum[table-format=2.3e2]{1.699e-3} & \tablenum[table-format=2.3e2]{1.537e-3}& \tablenum[table-format=2.4]{0.1652} & \tablenum[table-format=2.4]{0.1675}& \tablenum[table-format=2.4]{0.1674}\\
	$\rho_T(3)$& \tablenum[table-format=2.3e2]{1.755e-3} & \tablenum[table-format=2.3e2]{5.819e-4} & \tablenum[table-format=2.3e2]{3.943e-4}& \tablenum[table-format=2.4]{0.1211} & \tablenum[table-format=2.4]{0.1251}& \tablenum[table-format=2.4]{0.1250}\\
	\hline
	 $P(B=1)$& \tablenum[table-format=2.4]{0.3603} & \tablenum[table-format=2.4]{0.3603}  & \tablenum[table-format=2.4]{0.3577}& \tablenum[table-format=2.4]{0.5818} & \tablenum[table-format=2.4]{0.5818}  & \tablenum[table-format=2.4]{0.4642}\\
	$P(B=2)$& \tablenum[table-format=2.4]{0.6397} & \tablenum[table-format=2.4]{0.6397}  & \tablenum[table-format=2.4]{0.6423}& \tablenum[table-format=2.4]{0,1437} & \tablenum[table-format=2.4]{0,1437}& \tablenum[table-format=2.4]{0,2038}\\
	 $P(B=3)$&-&-&-& \tablenum[table-format=2.4]{0.2611} & \tablenum[table-format=2.4]{0.2611} & \tablenum[table-format=2.4]{0.3319}\\
	$P(B=4)$&-&-&-& \tablenum[table-format=2.4]{0.0135} & \tablenum[table-format=2.4]{0.0135}&  \tablenum[table-format=2.3e2]{6.932e-05}\\
	\hline
	$\begin{array}{c} running\\ time\end{array}$
	&-
	&\tablenum[table-format=2.2]{21.59}
	&\tablenum[table-format=2.2]{19.55}
	&-
	&\tablenum[table-format=2.2]{37.04}
	&\tablenum[table-format=2.2]{33.13}\\
	\hline
	\hline
	\end{tabular}}
\vspace{0.05cm} 
\centering 
\caption{Comparisson between the estimated descriptors via the $\bmmpp)$ and the $\mmpp$ with i.i.d. batches.} \label{ta:comp_bmmppvsiid}
{\footnotesize
	\parbox{6.2in}{
		\medskip
		\begin{center}
		\end{center}}}
\end{table}
\endgroup

\textcolor{black}{\subsection{Comparison with the EM algorithm and estimation of the $\bmmpp_2(K)/M/1$ queue}\label{EM}
This section serves two purposes. First, as commented in Section 1, some authors have considered inference for the general $\bmap$s, as \cite{Breuer} and \cite{Klemm} which adapt the EM algorithm for the $\bmap$. Therefore, one of the aims of this section is to compare the performance of the novel sequential fitting methods with that of the EM algorithm as implemented in \cite{Breuer}. Second, one of the main applications of $\bmap$ processes are related to queueing theory, see for example \cite{Lucantoni90,Ramaswami90,Luca91,Luca93,Lucantoni.transient} which explore theoretical properties of the $\bmap/G/1$ queueing system. In this section, we consider estimation for the $\bmmpp_2(K)/M/1$ queueing system where the $\bmmpp_2(K)$ is the arrival process in a single-server, first in first out queueing system with independent, Markovian service times. In particular, the inference approach described in Section \ref{fitting} will be combined with techniques from the queueing literature in order to estimate the stationary queue length distribution at departures.}

\textcolor{black}{In \cite{Breuer} and \cite{Klemm} the EM algorithm is considered and adapted for the $\bmap$. In order to compare the performance of the novel method with that of the EM algorithm, we consider the first simulated trace from Section \ref{simu}, with generator process and theoretical moments as in the second column of Table \ref{ta:comp_emvsmom}. To explore in depth the performance of the EM algorithm, two different starting points are considered; the first quite close to the true solution (forth column of Table \ref{ta:comp_emvsmom}) and a second point that is far away from the true solution (seventh column of Table \ref{ta:comp_emvsmom}). From the results in the table, some conclusions can be obtained. First, the estimated rate matrices provided by the EM algorithms seems more dependent on the starting solution than those under the novel approach; while the solutions obtained with the moments matching method are similar under the two choices of the initial values, the solutions given by the EM differ among them, being the first one more accurate than the second one. Concerning the estimation of the empirical moments, both methods provide similar values, all close to the empirical ones. Something similar occurs with respect to the log-likelihood values given the estimated parameters (second-to-last row of Table \ref{ta:comp_emvsmom}). Finally, concerning the running times, the EM algorithm turns out notably slower than the novel method, especially when the starting solution is not close to the true one.}

\begingroup 
\begin{table}[!h]
\sisetup{ 
	detect-all,
	table-number-alignment = center,
	table-figures-integer = 1,
	table-figures-decimal = 3,
	explicit-sign
} 
\resizebox{1.00\textwidth}{!}{
	\begin{tabular}{|c|c|c|c|c|c|c|c|c|}
	\hline
	& &&\multicolumn{3}{c|}{Close starting point}&\multicolumn{3}{c|}{Distant starting point}\\
	\hline
	&  \multicolumn{1}{c|}{$\begin{array}{c} Generator\\ Process\end{array}$}&  \multicolumn{1}{c|}{\emph{Empirical}}&\multicolumn{1}{c|}{$\begin{array}{c} Starting\\ solution\end{array}$} &   \multicolumn{1}{c|}{\emph{EM}} &  \multicolumn{1}{c|}{$\begin{array}{c} Sequential\\ approach\end{array}$}  &\multicolumn{1}{c|}{$\begin{array}{c} Starting\\ solution\end{array}$} & \multicolumn{1}{c|}{\emph{EM}} &  \multicolumn{1}{c|}{$\begin{array}{c} Sequential\\ approach\end{array}$}  \\ 
	\hline
	&&&&&&&&\\
	$\boldsymbol{D_0}$&$\left( \begin{array}{cc} \tablenum[table-format=2]{-5} & \tablenum[table-format=3]{2}  \\ \tablenum[table-format=2]{5} & \tablenum[table-format=3]{-10} \end{array}\right)$&
	-&
	$\left( \begin{array}{cc}\tablenum[table-format=3]{-10} &
	\tablenum[table-format=3]{3}\\ \tablenum[table-format=3]{5} & \tablenum[table-format=3]{-15} \end{array}\right)$&
	$\left( \begin{array}{cc}\tablenum[table-format=2.2]{-5.05} &
	\tablenum[table-format=3.2]{2.22}\\  \tablenum[table-format=2.2]{6.80} & \tablenum[table-format=3.2]{-12.07} \end{array}\right)$&  
	$\left( \begin{array}{cc}\tablenum[table-format=2.2]{-4.15} &
	\tablenum[table-format=3.2]{0.89}\\  \tablenum[table-format=2.2]{6.49} & \tablenum[table-format=3.2]{-12.24} \end{array}\right)$&
	 $\left( \begin{array}{cc} \tablenum[table-format=3]{-25} & \tablenum[table-format=3]{10}\\
	 \tablenum[table-format=3]{13} & \tablenum[table-format=3]{-27} \end{array}\right)$&
	$\left( \begin{array}{cc} \tablenum[table-format=3.2]{-11.51} & \tablenum[table-format=3.2]{8.42}\\
	\tablenum[table-format=2.2]{16.55} & \tablenum[table-format=3.2]{-20.61} \end{array}\right)$&
	$\left( \begin{array}{cc}	\tablenum[table-format=3.2]{-4.21} & \tablenum[table-format=3.2]{0.98}\\
	\tablenum[table-format=2.2]{4.54} & \tablenum[table-format=3.2]{-9.60}\end{array}\right)$\\
	&&&&&&&&\\
	$\boldsymbol{D_1}$&
	$\left( \begin{array}{cc}\tablenum[table-format=2]{1}&
	\tablenum[table-format=3]{0}\\
	\tablenum[table-format=2]{0}&
	\tablenum[table-format=3]{2}\end{array}\right)$&
	-&
	$\left( \begin{array}{cc}\tablenum[table-format=3]{4} &
	\tablenum[table-format=3]{0}\\ 
	\tablenum[table-format=3]{0} &
	\tablenum[table-format=3]{4}\end{array}\right)$& 
	$\left(\begin{array}{cc}\tablenum[table-format=2.2]{0.73} & \tablenum[table-format=3.2]{0} \\
	\tablenum[table-format=2.2]{0} & \tablenum[table-format=3.2]{3.02}\end{array}\right)$&
	$\left(\begin{array}{cc}\tablenum[table-format=2.2]{1.15} & \tablenum[table-format=3.2]{0} \\
	\tablenum[table-format=2.2]{0} & \tablenum[table-format=3.2]{2.77}\end{array}\right)$&
	$\left(\begin{array}{cc}\tablenum[table-format=3]{9} & \tablenum[table-format=3]{0}\\
	\tablenum[table-format=3]{0} & \tablenum[table-format=3]{6}\end{array}\right)$&
	$\left(\begin{array}{cc}\tablenum[table-format=2.2]{0.47} & \tablenum[table-format=3.2]{0}\\
	\tablenum[table-format=2.2]{0} & \tablenum[table-format=3.2]{2.85}\end{array}\right)$&
	$\left(\begin{array}{cc}\tablenum[table-format=2.2]{1.27} & \tablenum[table-format=3.2]{0}\\
	\tablenum[table-format=2.2]{0} & \tablenum[table-format=3.2]{1.63}\end{array}\right)$\\
	&&&&&&&&\\
	$\boldsymbol{D_2}$&
	$\left( \begin{array}{cc}\tablenum[table-format=2]{2}&
	\tablenum[table-format=3]{0}\\
	\tablenum[table-format=2]{0}&
	\tablenum[table-format=3]{3}\end{array}\right)$&
	-&
	$\left( \begin{array}{cc}\tablenum[table-format=3]{3} &
	\tablenum[table-format=3]{0} \\
	\tablenum[table-format=3]{0} &
	\tablenum[table-format=3]{6}\end{array}\right)$&
	$\left( \begin{array}{cc} \tablenum[table-format=2.2]{2.09} & \tablenum[table-format=3.2]{0}\\
	\tablenum[table-format=2.2]{0} & \tablenum[table-format=3.2]{2.25}\end{array}\right)$&
	$\left( \begin{array}{cc} \tablenum[table-format=2.2]{2.10} & \tablenum[table-format=3]{0}\\
	 \tablenum[table-format=2.2]{0} & \tablenum[table-format=3.2]{2.98}\end{array}\right)$&
	 $\left( \begin{array}{cc}\tablenum[table-format=3]{5} & \tablenum[table-format=3]{0}\\
	 \tablenum[table-format=3]{0} & \tablenum[table-format=3]{4}\end{array}\right)$&
	$\left(\begin{array}{cc}\tablenum[table-format=2.2]{2.62}& \tablenum[table-format=3.2]{0}\\
	\tablenum[table-format=2.2]{0} & \tablenum[table-format=3.2]{1.21}\end{array}\right)$&
	$\left(\begin{array}{cc}\tablenum[table-format=2.2]{1.96} & \tablenum[table-format=3.2]{0}\\
	\tablenum[table-format=2.2]{0} & \tablenum[table-format=3.2]{3.43}\end{array}\right)$\\
	&&&&&&&&\\
	$\mu_1$&\multicolumn{1}{c|}{0.28}&
	\multicolumn{1}{c|}{0.28}&
	\multicolumn{1}{c|}{-}&
	\multicolumn{1}{c|}{0.29}&
	\multicolumn{1}{c|}{0.28}&
	\multicolumn{1}{c|}{-}&
	\multicolumn{1}{c|}{0.29}&
	\multicolumn{1}{c|}{0.28}\\
	$\mu_2$&\multicolumn{1}{c|}{0.16}&
	\multicolumn{1}{c|}{0.16}&
	\multicolumn{1}{c|}{-}&
	\multicolumn{1}{c|}{0.17}&
	\multicolumn{1}{c|}{0.16}&
	\multicolumn{1}{c|}{-}&
	\multicolumn{1}{c|}{0.17}&
	\multicolumn{1}{c|}{0.16}\\
	$\mu_3$&\multicolumn{1}{c|}{0.138}&
	\multicolumn{1}{c|}{0.137}&
	\multicolumn{1}{c|}{-}&
	\multicolumn{1}{c|}{0.158}&
	\multicolumn{1}{c|}{0.138}&
	\multicolumn{1}{c|}{-}&
	\multicolumn{1}{c|}{0.148}&
	\multicolumn{1}{c|}{0.138}\\
	$\rho_T(1)$&\multicolumn{1}{c|}{\tablenum[table-format=1.2e2]{7.35e-3}}&
	\multicolumn{1}{c|}{\tablenum[table-format=1.2e2]{6.01e-3}}&
	\multicolumn{1}{c|}{-}&
	\multicolumn{1}{c|}{\tablenum[table-format=1.2e2]{2.92e-2}}&
	\multicolumn{1}{c|}{\tablenum[table-format=1.2e2]{5.99e-3}}&
	\multicolumn{1}{c|}{-}&
	\multicolumn{1}{c|}{\tablenum[table-format=1.2e2]{1.16e-1}}&
	\multicolumn{1}{c|}{\tablenum[table-format=1.2e2]{6.01e-3}}\\
	$\beta_1$&\multicolumn{1}{c|}{1.64}&
	\multicolumn{1}{c|}{1.62}&
	\multicolumn{1}{c|}{-}&
	\multicolumn{1}{c|}{1.67}&
	\multicolumn{1}{c|}{1.64}&
	\multicolumn{1}{c|}{-}&
	\multicolumn{1}{c|}{1.82}&
	\multicolumn{1}{c|}{1.62}\\
	$\eta_1$&\multicolumn{1}{c|}{0.46}&
	\multicolumn{1}{c|}{0.46}&
	\multicolumn{1}{c|}{-}&
	\multicolumn{1}{c|}{0.49}&
	\multicolumn{1}{c|}{0.46}&
	\multicolumn{1}{c|}{-}&
	\multicolumn{1}{c|}{0.53}&
	\multicolumn{1}{c|}{0.46}\\
	&&&&&&&&\\
	$l$&\multicolumn{1}{c|}{-}&\multicolumn{1}{c|}{-}&\multicolumn{1}{c|}{-270.66}&
	\multicolumn{1}{c|}{-114.37}&
	\multicolumn{1}{c|}{-114.63}&
	\multicolumn{1}{c|}{-414.25}&
	\multicolumn{1}{c|}{-114.10}&
	\multicolumn{1}{c|}{-114.70}\\
	&&&&&&&&\\
	$\begin{array}{c} running\\ time\end{array}$
	&\multicolumn{1}{c|}{-}&\multicolumn{1}{c|}{-}&\multicolumn{1}{c|}{-}&
	\multicolumn{1}{c|}{52.21}&
	\multicolumn{1}{c|}{0.23}&
	\multicolumn{1}{c|}{-}&
	\multicolumn{1}{c|}{98.72}&
	\multicolumn{1}{c|}{0.59}\\
	\hline
	\hline
	\end{tabular}}
\vspace{0.05cm} 
\centering 
\caption{Comparison between the EM algorithm and the novel sequential approach under two different starting solutions.} \label{ta:comp_emvsmom}
{\footnotesize
	\parbox{6.2in}{
		\medskip
		\begin{center}
		\end{center}}}
\end{table}
\endgroup

\begin{figure}%
\centering
\subfloat[Close starting point, $\rho=0.3$]{{\includegraphics[width=5cm]{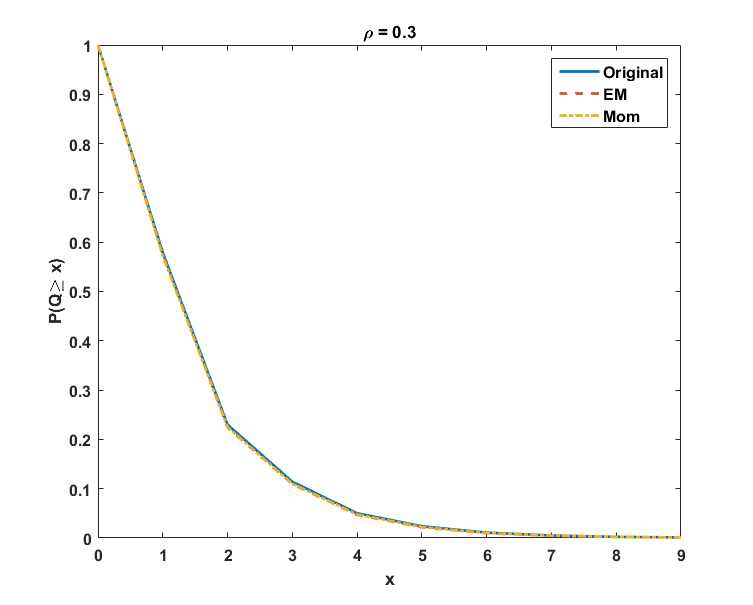}} }
\subfloat[Close starting point, $\rho=0.7$]{{\includegraphics[width=5cm]{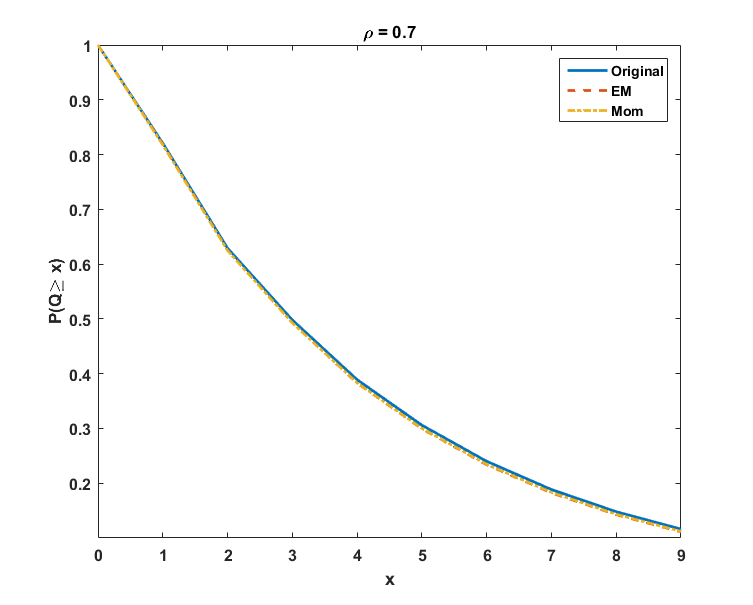} }}%

\subfloat[Distant starting point, $\rho=0.3$]{{\includegraphics[width=5cm]{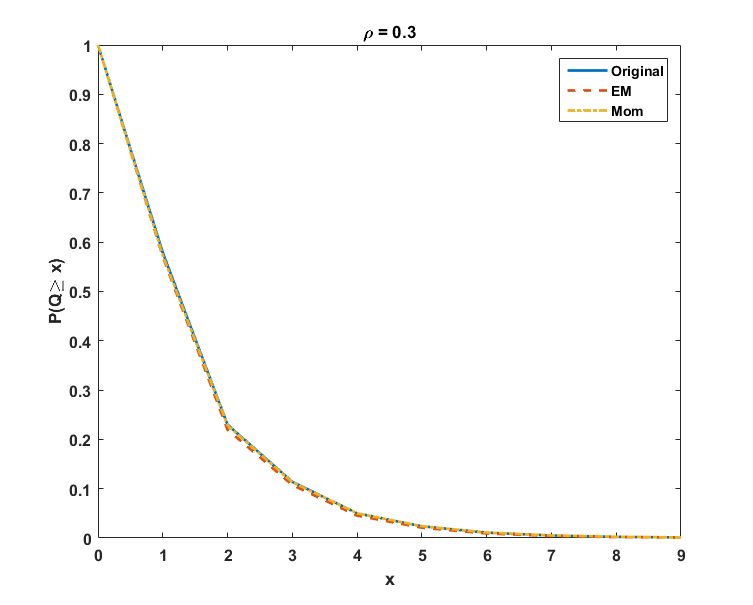} }}%
\subfloat[Distant starting point, $\rho=0.7$]{{\includegraphics[width=5cm]{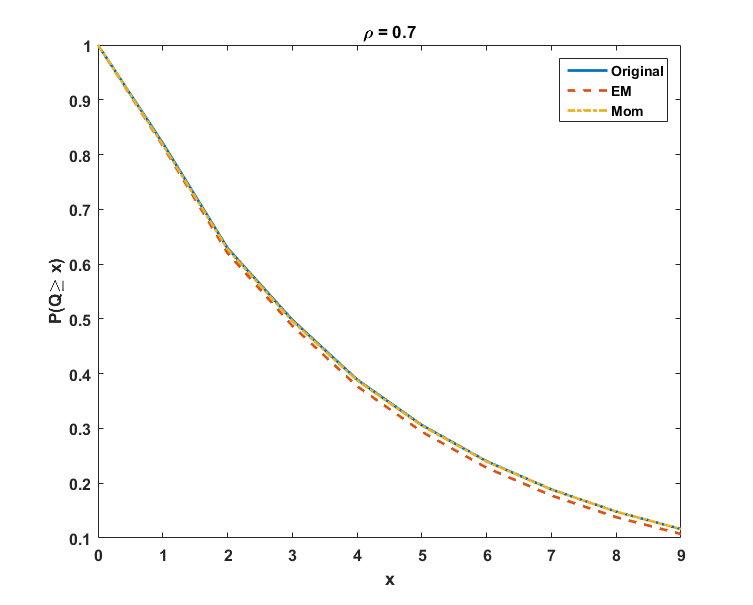} }}%
\caption{Estimated tail distributions of the queue length at departures in a $\bmmpp_2(2)/M/1$ queue.}\label{fig:Queue}%
\end{figure}

\textcolor{black}{Consider next the $\bmmpp_2(K)/M/1$ queueing system and denote by $\mu^\star<\infty$ the expected value of the service time. Then, the traffic intensity of this system is given by
\begin{equation*}\label{rho}
\rho=\lambda^\star/\mu^\star,
\end{equation*} where $\lambda^\star$ is the stationary arrival rate (inverse of the expected inter-event time), defined as
$$
\lambda^\star = 1/\mu_1,
$$
where $\mu_1$ is defined as in (\ref{eq:MAP-mu}). Now define $Z(t)$ to be the number of customers in the system (including in service, if any) at time $t$ and let $\tau_k$ be the epoch of the $k$-th departure from the queue, with $\tau_0=0$. If the system is stable ($\rho<1$), then for $i\geq 1$
\begin{equation*}
z_{i}=\lim_{k\rightarrow \infty}P\left[Z(\tau _{k})=i\right],
\end{equation*}
represents the stationary probability that the queue length is equal to $i$  when a departure occurs. Closed-form expressions for the generating function of the queue length distributions can be found in \cite{Luca93}. Assume that the simulated trace of inter-event times used in Table \ref{ta:comp_emvsmom} represents the inter-arrival times in a $\bmmpp_2(2)/M/1$ queue. Then, given the point estimates of the $\bmmpp_2(2)$ from the table, the numerical routines described in \cite{Luca93}, as well as in \cite{Abate} can be implemented to invert the generating function of of the queue length distribution. Figure \ref{fig:Queue} depicts the estimated tail distributions of the queue length at departures for two different services times (that is, for two different traffic intensities, $\rho=0.3,\ 0.7$) and  for both  solutions (from the sequential fitting approach and EM algorithm) and under the two possible choices of starting points considered in Table \ref{ta:comp_emvsmom}. In the figure, the solid line represents the true distribution, the dashed line is the estimated function using the EM solution and finally, the dotted line depicts the estimated tail distribution under the solution obtained by the sequential fitting method. From the figure some comments can be made. First, as it is expected, larger values for the tail distribution are obtained in the case of $\rho=0.7$, a  consequence of the higher degree of saturation of the system.  Second, an additional expected fact is that the estimated tail distributions using a close starting point are slightly  more accurate than those obtained under the distant starting points. Finally, in the case of the distant starting point with $\rho=0.7$, the sequential fitting approach leads to a slightly more precise solutions than the EM algorithm.}

\subsection{Numerical illustration on a teletraffic real data set}\label{teletraffic}

In this section we illustrate the performance of the novel approach for fitting a well-referenced  database, namely the Bellcore Aug89 data set, which has been considered in a number of papers concerning teletraffic modeling, see \textcolor{black}{\cite{Horv05}},\cite{Ram08,PepadPlN,Li10,Kriege11,oka11,RodInf,Cas16}.

\subsubsection*{Data description}
\begin{figure}[h!]
\begin{center}
\begin{tabular}{c c}
\includegraphics[height=1.7in]{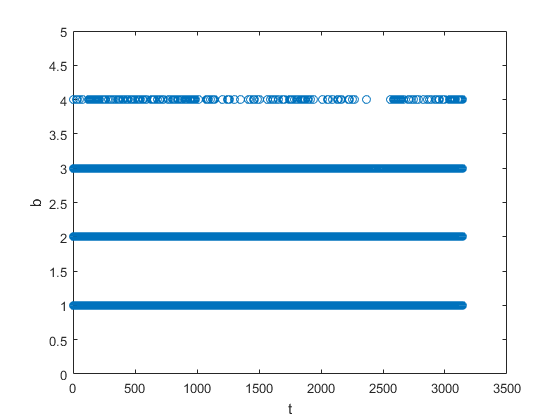}&
\includegraphics[height=1.7in]{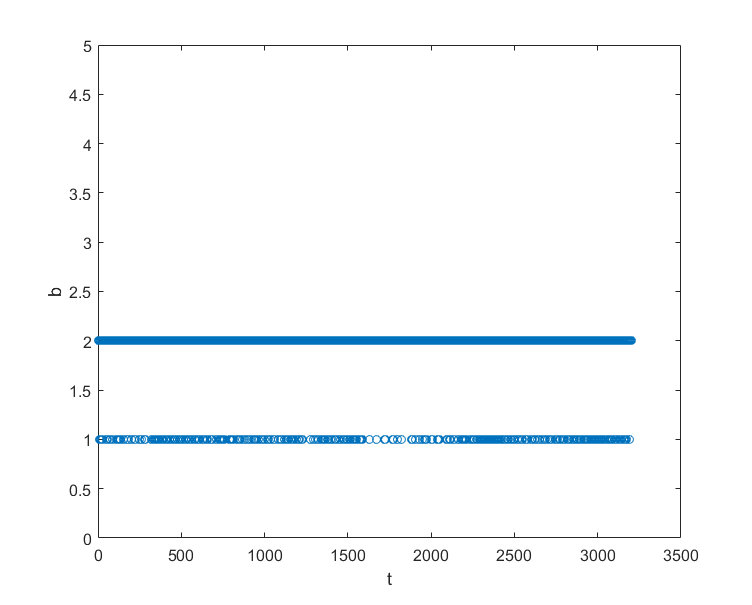}
\end{tabular}
\end{center}
\caption{{Left panel: packets arriving in batches of sizes $1,\ldots,4$, observed in intervals of length $10^{-3}$ seconds (Data set in format I). Right panel: packets divided into \emph{small} (batch size equal to 1) and \emph{large} (batch size equal to 2) (Data set in format II).}}\label{fig:realdata}
\end{figure}

The data set BC-pAug89, available at the web site\\
\begin{center}
 {\tt http://ita.ee.lbl.gov/html/contrib/BC.html}
 \end{center}
 consists of one million of packet arrivals seen on Ethernet at the Bellcore Morristown Research and Engineering facility. The trace began at 11:25 on August 29, 1989, and ran for about $3142.82$ seconds until the arrival of one million packets (of different size each). The times are originally expressed to 6 places after the decimal point (milisecond resolution), which implies that packets arrive in isolated way. However, if instead of observing the process every $10^{-6}$ seconds, it is observed every $10^{-3}$ seconds, then this form of aggregation leads to packets arriving in batches, with batches sizes varying from $1$ to $4$, as shown by the left panel of Figure \ref{fig:realdata}. \textcolor{black}{This structure of the data set will be called from now on Data set in format I. On the other hand, the original data set can be viewed from a different perspective if the size of the packets is taken into account. Due to the Ethernet protocol the size of the packets takes $866$ different values ranging from $64$ to $1518$ bytes. Therefore, packets can be divided into \emph{small} packets, when the size is lower than $100$ bits, and \emph{large}, otherwise, see the right panel of Figure \ref{fig:realdata}. This new format, proposed in analogous way in \cite{Klemm}, will be called henceforth Data set in format II, where the batch size equal to $1$ will refer to \emph{small} sizes, and a batch size of $2$ will be used to refer to the \emph{large} sizes.}  

\begin{figure}[h!]
\begin{center}
\begin{tabular}{c}
\includegraphics[height=1.7in]{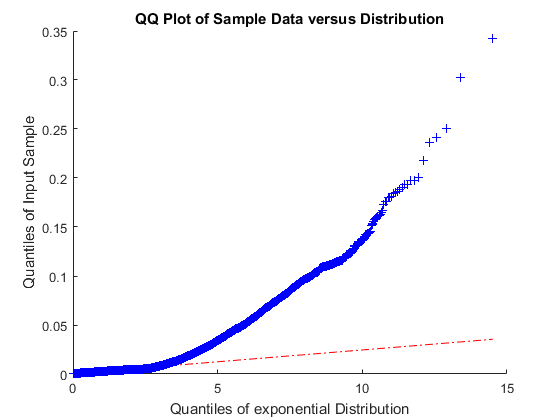}
\end{tabular}
\end{center}
\caption{Empirical quantiles of the inter-arrival times of Data set in format I versus those of a fitted exponential distribution.}\label{fig:qqplot2}
\end{figure}

\textcolor{black}{Consider first Data set in format I.  There are strong reasons to not assume a Poisson process for the Data set in format I . First, the  average, median, variation coefficient, minimum and maximum value of the inter-arrivals times are $0.0036$, $0.0020$, $1.6553$, $1\times 10^{-3}$ and $0.3420$ seconds, respectively, which suggests a right-skewed distribution with a tail longer than that of an exponential distribution. Indeed, Figure \ref{fig:qqplot2} shows the empirical quantiles comparison to that of the fitted (via MLE) exponential distribution. Note how the larger empirical quantiles are far from the fitted ones. Something similar occurs with Data set in format II, where the average, median, variation coefficient, minimum and maximum value of the inter-arrivals times are given by $0.0031$, $0.0020$, $1.7954$, $2\times 10^{-5}$ and $0.3419$ seconds. In addition, the empirical first-lag correlation coefficients of the inter-arrival times are $0.1908$ and $0.2$, respectively. This implies that a model capturing dependence between the arrivals may turn out suitable. Since arrivals occur in batches, a $\bmmpp_2(2)$ and a $\bmmpp_2(2)$ will be fitted to the data sets using the novel sequential fitting approach; the results shall be shown in the next section.}

\subsubsection*{Results}
\textcolor{black}{The sequential algorithm described in Section \ref{fitting} is applied to fit the teletraffic data sets. Table \ref{ta:comp_empvsest} shows the empirical values of a set of descriptors concerning the inter-arrivals times distribution, the batch sizes distribution and joint moments, as well as the estimated values under a $\bmmpp_2(4)$ and $\bmmpp_2(2)$ models for Data set in format I and II, respectively. From the first to the $10$-th row, the fitted values to the characterizing moments, according to Theorem 2, are provided. Then, the estimated coefficient of variation, skewness and kurtosis are shown. The $14$-th and $15$-th rows concern the first and second moments of the batch size. Then, some descriptors related to the correlation between the inter-arrival times and the batches and the autocorrelation coefficients of the inter-event times are also depicted. The probability mass distribution of the batch size is also shown. Most of the quantities are well estimated by the considered models, with the exception of the values of $\rho_T(2)$ and $\rho_T(3)$ which are slightly underestimated. }

\begin{figure}[h!]
\begin{center}
\begin{tabular}{c c}
\includegraphics[height=1.7in]{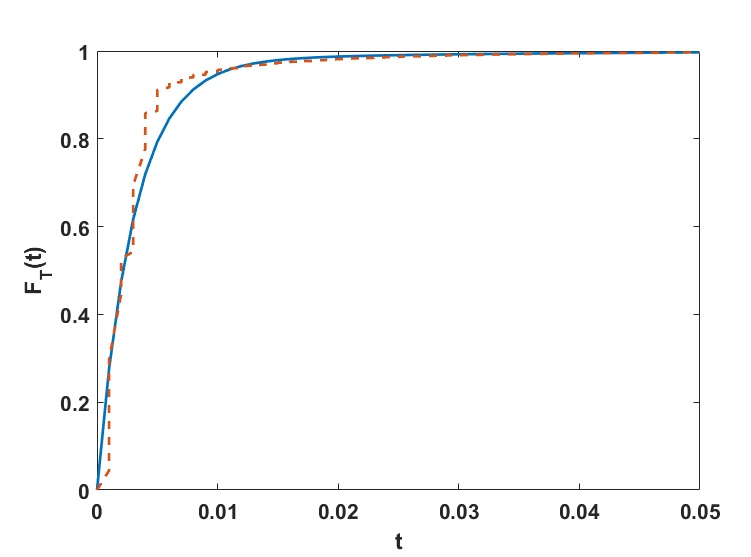}&
\includegraphics[height=1.7in]{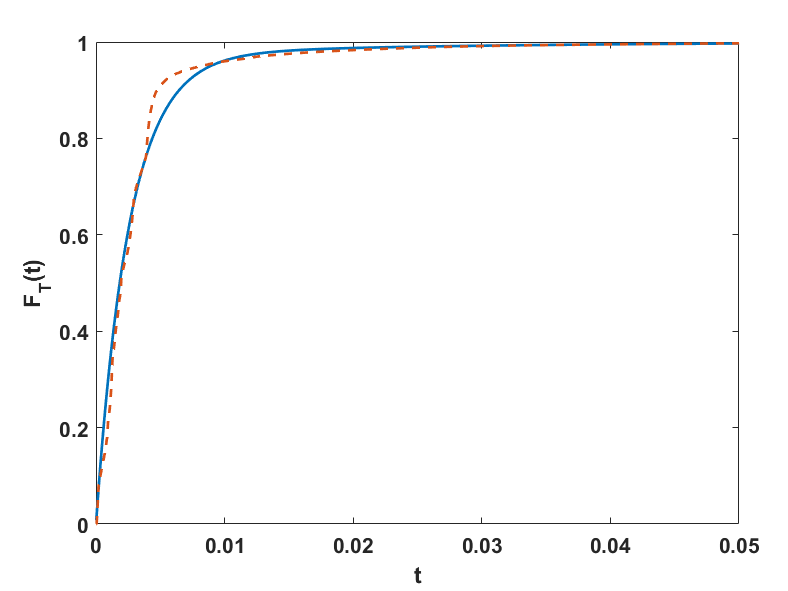}
\end{tabular}
\end{center}
\caption{Left panel: Estimated cdf (dashed line) under the $\bmmpp_2(2)$ versus the empirical cdf (solid line) of the inter-arrival times. Right panel: Estimated cdf (dashed line) under the $\bmmpp_2(4)$ versus the empirical cdf (solid line) of the inter-arrival time.}\label{fig:ft1}
\end{figure}

\textcolor{black}{Table \ref{ta:comp_empvsest} also shows a comparisson between  the general model proposed in the paper and a $\mmpp$ with i.i.d batches. For the estimation of the $\mmpp$s with i.i.d batches, sequential algorithm described in Section \ref{fitting} was used, adding $q_k(\hat{y}-\hat{x})=w_k(\hat{r}-\hat{u})$ as a restriction to each $(P_k)$ optimization problem. In Table \ref{ta:comp_empvsest} can be appreciated that in the case of Data set in format I, for which $\rho(T, B)$ and $\rho(B)$ are almost null, both estimations are quite similar. But by slightly increasing these amounts  for Data set in format II the estimation using the general $\bmmpp$ improves over the case with independent bathes. In conclusion, although the  correlations or autocorrelations shown by the data are negligible, it is more reliable to adjust the general model since the computation time does not increase substantially but it does improve the quality of the adjustment in general. The good performance of the fitted models is also supported by Figure \ref{fig:ft1} which depicts the fit to the empirical distribution functions of the inter-arrival times.}

\begingroup 
\begin{table}[!h]
\sisetup{ 
            detect-all,
            table-number-alignment = center,
            table-figures-integer = 1,
            table-figures-decimal = 3,
            explicit-sign
} 
\resizebox{0.95\textwidth}{!}{%
\begin{tabular}{r|ccc|ccc}
\hline
  & \multicolumn{3}{c|}{Data set in format I} & \multicolumn{3}{c}{Data set in format II} \\ 
 & \multirow{ 2}{*}{Emp} & \multirow{ 2}{*}{Est BMMPP} & Est MMPP with &\multirow{ 2}{*}{Emp} & \multirow{ 2}{*}{Est BMMPP} & Est MMPP with \\ 
 &  & & i.i.d. batches  & & & i.i.d. batches \\
\hline
 $\mu_1$& \tablenum[table-format=1.4e2]{3.5625E-03} & \tablenum[table-format=1.4e2]{3.5625E-03} & \tablenum[table-format=1.4e2]{3.5625E-03}
& \tablenum[table-format=1.4e2]{3.1428E-03} & \tablenum[table-format=1.4e2]{3.1428E-03}& \tablenum[table-format=1.4e2]{3.1428E-03}\\
$\mu_2$& \tablenum[table-format=1.4e2]{4.7465E-05} & \tablenum[table-format=1.4e2]{4.7465E-05} & \tablenum[table-format=1.4e2]{4.7465E-05}
& \tablenum[table-format=1.4e2]{4.1718E-05} & \tablenum[table-format=1.4e2]{2.0859E-05} & \tablenum[table-format=1.4e2]{4.1718E-05}\\
$\mu_3$& \tablenum[table-format=1.4e2]{2.2802E-06} & \tablenum[table-format=1.4e2]{2.2802E-06} & \tablenum[table-format=1.4e2]{2.2802E-06}
& \tablenum[table-format=1.4e2]{2.0104E-06} & \tablenum[table-format=1.4e2]{2.0104E-06}& \tablenum[table-format=1.4e2]{2.0104E-06}\\
$\rho_T(1)$& \tablenum[table-format=1.4e2]{0.1908} & \tablenum[table-format=1.4e2]{0.1908}& \tablenum[table-format=1.4e2]{0.1908}
& \tablenum[table-format=1.4e2]{0.2} & \tablenum[table-format=1.4e2]{0.2}& \tablenum[table-format=1.4e2]{0.2}\\
$\beta_1^{(1)}$& \tablenum[table-format=1.4e2]{1.1241} & \tablenum[table-format=1.4e2]{1.1096}& \tablenum[table-format=1.4e2]{1.1040}
& \tablenum[table-format=1.4e2]{1.8121} & \tablenum[table-format=1.4e2]{1.8121}& \tablenum[table-format=1.4e2]{1.7788}\\
$\eta^{(1)}$& \tablenum[table-format=1.4e2]{3.8663E-3} & \tablenum[table-format=1.4e2]{3.9189E-3}& \tablenum[table-format=1.4e2]{3.9329E-3}
& \tablenum[table-format=1.4e2]{5.4932E-3} & \tablenum[table-format=1.4e2]{5.4932E-3}& \tablenum[table-format=1.4e2]{5.5905E-3}\\
$\beta_1^{(2)}$& \tablenum[table-format=1.4e2]{1.8849} & \tablenum[table-format=1.4e2]{1.8970}& \tablenum[table-format=1.4e2]{1.9019}
& - & - &- \\
$\eta^{(2)}$& \tablenum[table-format=1.4e2]{6.8387E-3} & \tablenum[table-format=1.4e2]{6.7898E-3} & \tablenum[table-format=1.4e2]{6.7756E-3}
& - & - &- \\
$\beta_1^{(3)}$& \tablenum[table-format=1.4e2]{1.9915} & \tablenum[table-format=1.4e2]{1.9935}
& \tablenum[table-format=1.4e2]{1.9942}
& - & - &- \\
$\eta^{(3)}$& \tablenum[table-format=1.4e2]{7.1082E-3} & \tablenum[table-format=1.4e2]{7.1037E-3} & \tablenum[table-format=1.4e2]{7.1044E-3}
& - & - &-\\
\hline 
 $CV$& \tablenum[table-format=1.4e2]{1.6553} & \tablenum[table-format=1.4e2]{1.6553} & \tablenum[table-format=1.4e2]{1.6553}
& \tablenum[table-format=1.4e2]{1.7954} & \tablenum[table-format=1.4e2]{1.7954}& \tablenum[table-format=1.4e2]{1.7954}\\
$Skewness$& \tablenum[table-format=2.4e2]{11.1199} & \tablenum[table-format=2.4e2]{11.1199}& \tablenum[table-format=2.4e2]{11.1199}
& \tablenum[table-format=2.4e2]{11.1896} & \tablenum[table-format=2.4e2]{11.1896}& \tablenum[table-format=2.4e2]{11.1896}\\
 $Kurtosis$& \tablenum[table-format=3.4e2]{170.3031} & \tablenum[table-format=3.4e2]{168.1339}
& \tablenum[table-format=3.4e2]{168.1340}
& \tablenum[table-format=3.4e2]{179.2370} & \tablenum[table-format=3.4e2]{166.8824}& \tablenum[table-format=3.4e2]{166.8824}\\
\hline 
$\beta_1$& \tablenum[table-format=1.4e2]{1.1335} & \tablenum[table-format=1.4e2]{1.1162}& \tablenum[table-format=1.4e2]{1.1100}
& \tablenum[table-format=1.4e2]{1.8121} & \tablenum[table-format=1.4e2]{1.8121}& \tablenum[table-format=1.4e2]{1.7788}\\
 $\beta_2$& \tablenum[table-format=1.4e2]{1.4205} & \tablenum[table-format=1.4e2]{1.3618}  & \tablenum[table-format=1.4e2]{1.3424}
& \tablenum[table-format=1.4e2]{3.4364} & \tablenum[table-format=1.4e2]{3.4363} & \tablenum[table-format=1.4e2]{3.3365}\\
\hline
$Corr(T,B)$& \tablenum[table-format=1.4e2]{-0.0707} & \tablenum[table-format=1.4e2]{-0.0180} & \tablenum[table-format=1.4e2]{0} 
& \tablenum[table-format=1.4e2]{-0.0916} & \tablenum[table-format=1.4e2]{-0.0916} & \tablenum[table-format=1.4e2]{0}\\
$\rho_B(1)$& \tablenum[table-format=1.4e2]{0.0500} & \tablenum[table-format=1.4e2]{6.1083E-3}& \tablenum[table-format=1.4e2]{0}& \tablenum[table-format=1.4e2]{0.1037} & \tablenum[table-format=1.4e2]{0.1141} & \tablenum[table-format=1.4e2]{0}\\
$\rho_T(2)$& \tablenum[table-format=1.4e2]{0.1791} & \tablenum[table-format=1.4e2]{0.1146}& \tablenum[table-format=1.4e2]{0.1146}
& \tablenum[table-format=1.4e2]{0.1893} & \tablenum[table-format=1.4e2]{0.1160}& \tablenum[table-format=1.4e2]{0.1160}\\
$\rho_T(3)$& \tablenum[table-format=1.4e2]{0.1278} & \tablenum[table-format=1.4e2]{0.0689}& \tablenum[table-format=1.4e2]{0.0689}
& \tablenum[table-format=1.4e2]{0.1390} & \tablenum[table-format=1.4e2]{0.0673}& \tablenum[table-format=1.4e2]{0.0673}\\
\hline
 $P(B=1)$& \tablenum[table-format=1.4e2]{0.8759} & \tablenum[table-format=1.4e2]{0.8904}  & \tablenum[table-format=1.4e2]{0.8960}
& \tablenum[table-format=1.4e2]{0.1879} & \tablenum[table-format=1.4e2]{0.1879}& \tablenum[table-format=1.4e2]{0.2212}\\
$P(B=2)$& \tablenum[table-format=1.4e2]{0,1151} & \tablenum[table-format=1.4e2]{0,1031}& \tablenum[table-format=1.4e2]{0,0891}
& \tablenum[table-format=1.4e2]{0.8121} & \tablenum[table-format=1.4e2]{0.8121}& \tablenum[table-format=1.4e2]{0.7788}\\
 $P(B=3)$& \tablenum[table-format=1.4e2]{0.0085} & \tablenum[table-format=1.4e2]{0.0065} & \tablenum[table-format=1.4e2]{0.0058}
& - & - &-\\
$P(B=4)$& \tablenum[table-format=1.4e2]{4.7042e-4} & \tablenum[table-format=1.4e2]{1.7143e-05}&  \tablenum[table-format=1.4e2]{1.0262e-04}& - & - &-\\
\hline
\hline
\end{tabular}}
\vspace{0.05cm} 
\centering 
\caption{Empirical and estimated descriptors via the $\bmmpp_2(4)$ (Data set in format I) and the $\bmmpp_2(2)$ (Data set in format II).} \label{ta:comp_empvsest}
 {\footnotesize
\parbox{6.2in}{
\medskip
\begin{center}
\end{center}}}
\end{table}
\endgroup

\begin{figure}[h!]
\begin{center}
\begin{tabular}{c c}
\includegraphics[height=1.7in]{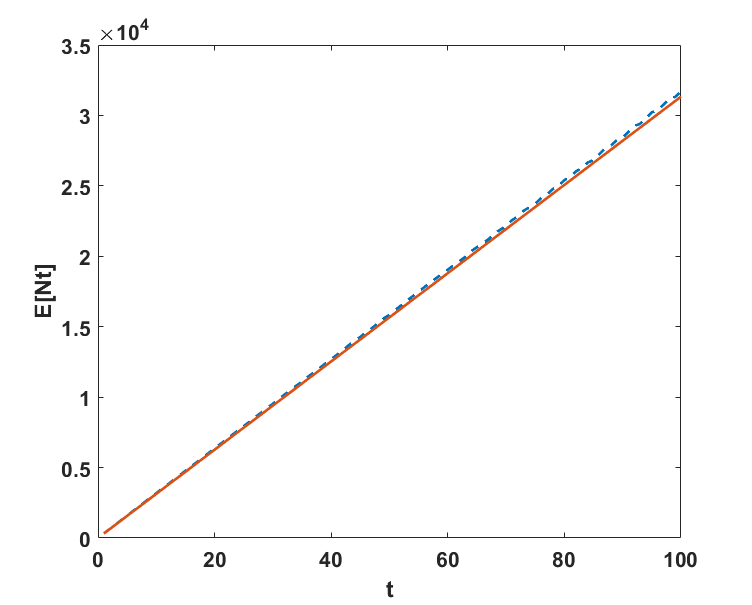}&
\includegraphics[height=1.7in]{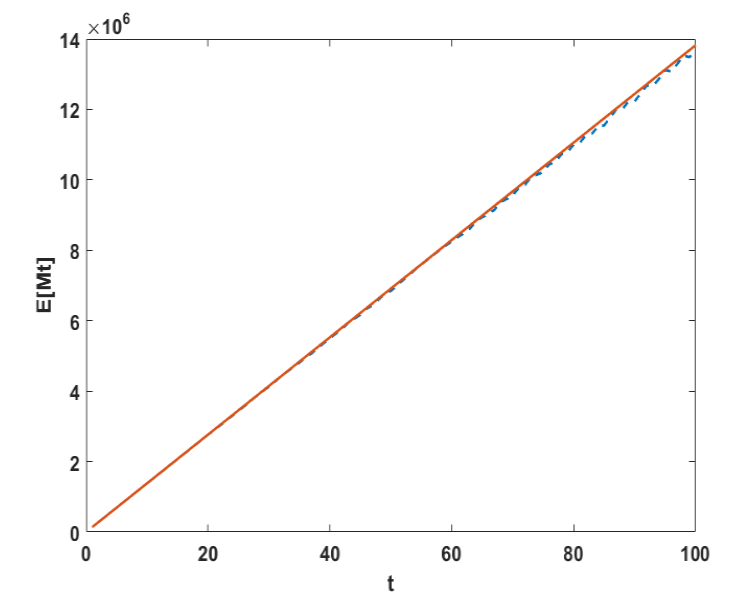}\\
\includegraphics[height=1.7in]{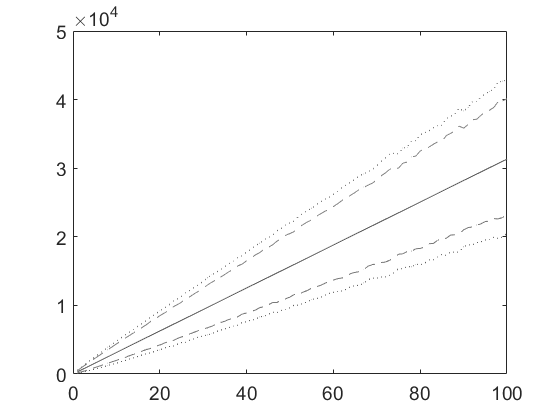}&
\includegraphics[height=1.7in]{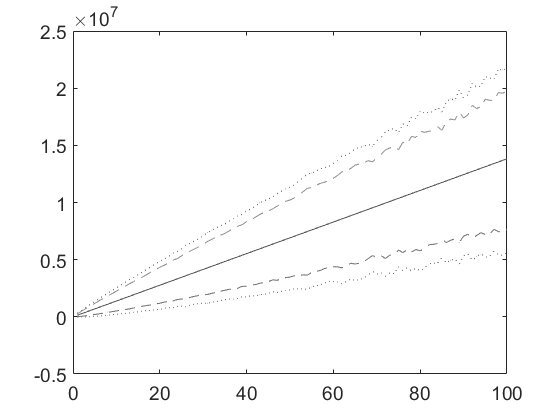}\\
\end{tabular}
\end{center}
\caption{Top panels: Estimated (dashed line) and empirical (solid line) expected number of arrivals, for Data sets in formats I and II, respectively. Bottom panels: Estimated $E[N(t)]$ (solid line) and $E[N(t)]\pm kSd(N(t))$, for $k=1$ (dashed line) and $k=2$ (dotted line), for Data sets in formats I and II, respectively.}\label{fig:ENt1}
\end{figure}

\begin{figure}[h!]
\begin{center}
\begin{tabular}{c c}
\includegraphics[height=1.7in]{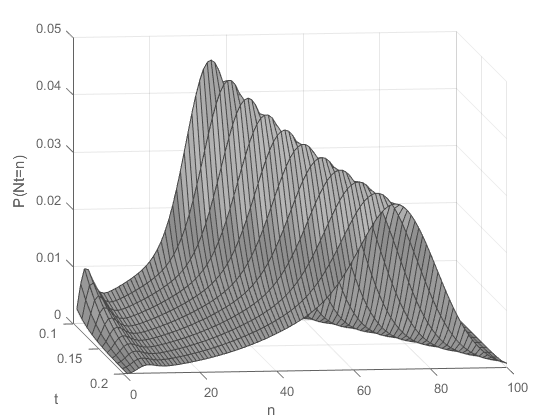}&
\includegraphics[height=1.7in]{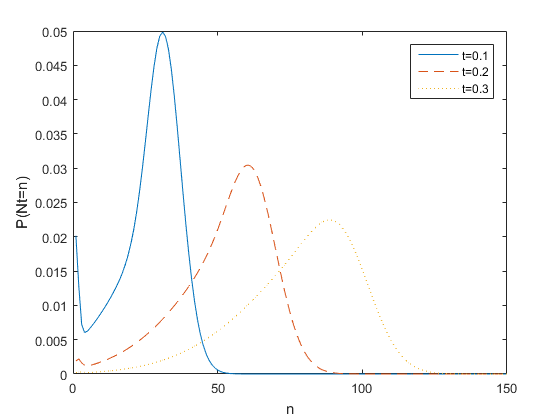}
\end{tabular}
\end{center}
\caption{Left panel: estimated density function of the number of arrivals in different time instants for Data set in format I. Right panel: distribution of the number of \emph{large} packets for Data set in format II for three different time instants.}\label{fig:Pnt_tran1}
\end{figure}

Next, we focus on some quantities of interest associated to the the counting process, see Section \ref{counting}. The top panels of Figure \ref{fig:ENt1} show the estimated and empirical expected number of arrivals in the interval $(0,100)$ for both Data set in formats I and II. The bottom panels depict the estimated intervals centered on $E[N(t)]\pm kSd(N(t))$, for $k=1,2$, where $Sd(N(t))$ denotes the standard deviation of the number of counts, computed from (\ref{variance_count}). On the other hand, Figure \ref{fig:Pnt_tran1} illustrates the estimated probabilities $p(n,t)$ as in (\ref{PN}). The left panel shows the estimated probabilities for Data set in format I for $n\in [0,100]$ and $t=[0.1,0.2]$. As it can be observed, the sequence of functions for different values of $t$ are bimodal, with a maximum around a high number of $n$, and another local maximum for a small value of $n$. In addition, the probability functions are not symmetric with a left tail that is longer than the right tail. \textcolor{black}{Concerning Data set in format II, the left panel of Figure \ref{fig:Pnt_tran1} shows the probabilities of the counts, for different time values and for \emph{large} sizes. It can be seen how the variability of the variable increases with the value of $t$.}

\begin{figure}[h!]
\begin{center}
\begin{tabular}{c}
\includegraphics[height=1.7in]{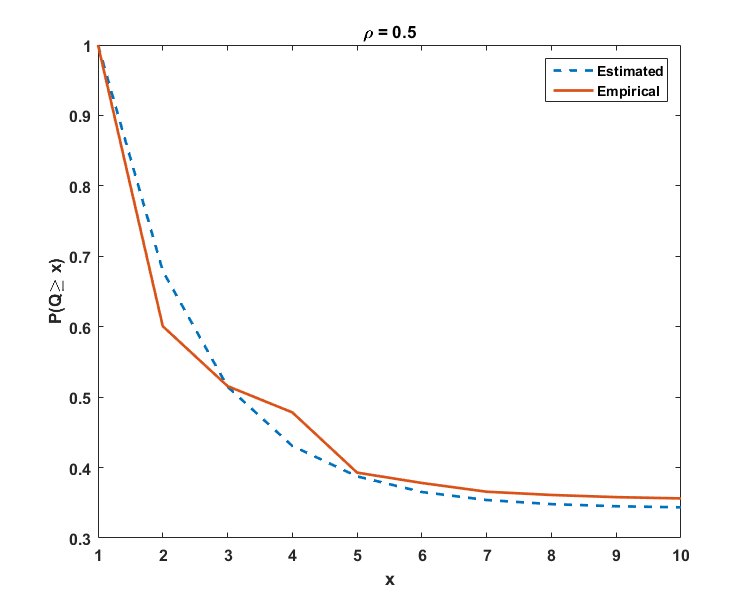}
\end{tabular}
\end{center}
\caption{Estimated tail distribution of the queue length at departures in a $\bmmpp_2(4)/M/1$ with $\rho=0.5$ assuming that the inter-arrival times of Data set in format I constitute the observed arrival process.}\label{fig:queue}
\end{figure}

\textcolor{black}{Finally, the queue length distribution of the $\bmmpp_2(4)/M/1$ queueing system was estimated, under the assumption that the inter-arrival times of Data set in format I constitute the observed arrival process. For that, a traffic intensity $\rho=0.5$ was set. Figure \ref{fig:queue} shows the resulting tail distribution.}

\begin{figure}[htb]
\centering
\includegraphics[width=7cm]{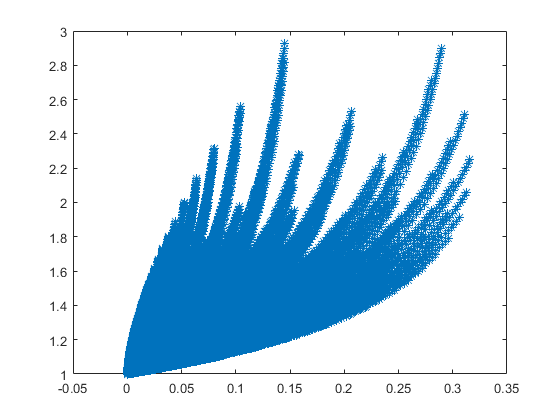}
\caption{Scatter plot of $\rho_T(1)$ versus the coefficient of variation of the inter-event times from $700000$ simulated $BMMPP$s}\label{fig:rhovscv}
\end{figure}

\section{Conclusions}\label{conclusions}

This paper considers the batch counterpart of the two-state Markov modulated Poisson Process. The point process, noted as $\bmmpp_2(K)$ turns out of interest in real-life contexts as reliability or queueing, since it allows for the modeling of dependent inter-event times and dependent batch sizes. The contribution of this paper is two-fold. On one hand, it is proven that the $\bmmpp_2(K)$, represented by $2(K+1)$ parameters, is completely characterized in terms of a set of $2(K+1)$ moments related to the inter-event time distribution as well as to the batch size distribution. On the other hand, an inference approach for fitting real data sets based on a moments matching method is described. The method involves solving, in iterative way, $K-1$ optimization problems with two unknowns, yielding an efficient and tractable algorithm. \textcolor{black}{The performance of the novel inference technique is illustrated using both simulated and a real teletraffic trace, for which the queue length distribution at departures in a $\bmmpp_2(K)/M/1$ queue are estimated. The method is also compared to the classic EM algorithm which has been considered by previous works dealing with inference for the $\bmap$. The results show that the novel approach turns out faster and less dependent on starting points than the EM algorithm.}

\textcolor{black}{Prospects regarding this work concern both applied and theoretical issues. In the first case, given that higher order $\bmmpp_m(K)$ are expected to show more versatility for modeling purposes \citep{Rod161}, it is of interest to develop inference methods in these cases. A moment-matching approach similar to the proposed in this paper could be considered for higher order $\bmmpp$s (which are known to be identifiable, \cite{Yera}). However, the set of moments characterizing $\bmmpp_m(K)$ processes is still unknown when $m\geq 3$. From a theoretical viewpoint, a challenging problem to be considered is related to the correlation structures (of both inter-event times and batch sizes) of the $\bmmpp_m(K)$, for $m\geq 3$. Similar approaches as in \cite{RamirezCoboyCarrizosa2012,Rod161} shall be taken into account to address this issue. Finally, another theoretical problem that needs to be examined in more detail refers to the samples sizes required for the estimation method. In this direction,  \cite{BayesPepa} suggest that the values of the coefficient of variation of the inter-event times ($CV$) and the first-lag autocorrelation coefficient ($\rho_T(1)$) are positively correlated. This would imply that the required sample size should increase with the value of the correlation between consecutive events. From Figure \ref{fig:rhovscv}, it can be seen that even though there exists processes for which the $CV$ is high and the value of $\rho_T(1)$ is low, it is true that high values of $\rho_T(1)$ seem to be linked to values of the $CV$ larger than a lower bound ($CV\sim 1.6$). On the contrary, if $\rho_T(1)$ very close to zero, then the $CV$ seems to closer to $1$. This problem is an open question that, together with the previous issues, will be undertaken in future work.}

\section*{Acknowledgments}
\textcolor{black}{The authors would like to thank the anonymous referees whose comments and suggestions improved the understanding of this paper.} Research partially supported by research grants and projects MTM2015-65915-R and ECO2015-66593-P (Ministerio de Econom\'ia y Competitividad, Spain) and P11-FQM-7603, FQM-329 (Junta de Andaluc\'ia, Spain) and Fundaci\'on BBVA.

\section*{Bibliography}
\bibliographystyle{myapalike}
\bibliography{ref_paper}

\end{document}